\documentclass[pdflatex,sn-mathphys-num]{sn-jnl}

\usepackage{graphicx}%
\usepackage{multirow}%
\usepackage{amsmath,amssymb,amsfonts}%
\usepackage{amsthm}%
\usepackage{mathrsfs}%
\usepackage[title]{appendix}%
\usepackage{xcolor}%
\usepackage{textcomp}%
\usepackage{manyfoot}%
\usepackage{booktabs}%
\usepackage{algorithm}%
\usepackage{algorithmicx}%
\usepackage{algpseudocode}%
\usepackage{listings}%
\usepackage{adjustbox}
\usepackage{fontawesome}
\usepackage{inconsolata} 
\usepackage[font=footnotesize,labelfont=bf]{caption}
\usepackage{soul}
\usepackage{hyperref}
\usepackage{lscape}
\usepackage{enumitem}
\usepackage{fontawesome}
\usepackage{multirow}
\usepackage[framemethod=tikz]{mdframed}
\usepackage{adjustbox}
\usepackage{float}
\usepackage{booktabs}
\usepackage{caption}

\usepackage{tikzsymbols}
\usepackage{textcomp}
\definecolor{diffstart}{named}{gray}
  \definecolor{diffincl}{named}{green}
  \definecolor{diffrem}{named}{orange}
\usepackage{xspace}
\usepackage[most]{tcolorbox}
\definecolor{codebg}{RGB}{255,255,255}
\definecolor{codeframe}{RGB}{180,180,180}
\definecolor{linenum}{RGB}{110,110,110}

\definecolor{diffRed}{RGB}{215,58,73}
\definecolor{diffGreen}{RGB}{128,128,128}

\lstdefinestyle{paperCode}{
  basicstyle=\ttfamily\footnotesize,
  numbers=left,
  numberstyle=\ttfamily\footnotesize\color{linenum},
  numbersep=3pt,
  xleftmargin=1.5em,      
  frame=none,
  showstringspaces=false,
  tabsize=2,
  breaklines=true,
  breakatwhitespace=true,
  escapeinside={(*@}{@*)}
}

\newtcblisting{codebox}[1][]{
  enhanced,
  listing only,
  listing options={
    style=paperCode,
    #1
  },
  colback=codebg,
  colframe=codeframe,
  boxrule=0.4pt,
  arc=3pt,
  left=0.5pt,
  right=0.5pt,
  top=1pt,
  bottom=1pt
}
\usepackage{textcomp}
  \definecolor{diffstart}{named}{gray}
  \definecolor{diffincl}{HTML}{00a67d} 
  \definecolor{diffrem}{named}{red}

 \lstdefinelanguage{diff}{
    basicstyle=\ttfamily\small,
    morecomment=[f][\color{diffstart}]{@@},
    morecomment=[f][\color{diffincl}]{+\ },
    morecomment=[f][\color{diffrem}]{-\ },
    }

\theoremstyle{thmstyleone}%
\theoremstyle{thmstyletwo}%

\theoremstyle{thmstylethree}%

\usepackage{listings}
  \lstdefinelanguage{diff}{
    basicstyle=\ttfamily\normalsize,
    morecomment=[f][\color{diffstart}]{@@},
    morecomment=[f][\color{diffincl}]{+\ },
    morecomment=[f][\color{diffrem}]{-\ },
    }

\lstdefinestyle{thrust1}{
  backgroundcolor=\color{white},
  basicstyle=\ttfamily\scriptsize,
  frame=single,
  breaklines=true,
  numbers=left,
  numberstyle=\tiny\color{gray},
  showstringspaces=false,
  moredelim=**[is][\color{red}]{@r@}{@},     
  moredelim=**[is][\color{green!50!black}]{@g@}{@}, 
  moredelim=**[is][\color{gray}]{@c@}{@},     
}

\lstdefinestyle{thrust1python}{
  language=Python,
  basicstyle=\ttfamily\scriptsize,
  keywordstyle=\color{red!80!black}\bfseries,
  commentstyle=\color{gray},
  stringstyle=\color{stringgray},
  identifierstyle=\color{black},
  emph={update, modify, _qs, _object_key, True, False, None},
  emphstyle=\color{keywordblue},
  showstringspaces=false,
  breaklines=true,
  backgroundcolor=\color{white},
  frame=single
}

\definecolor{diffadd}{rgb}{0.0,0.4,0.0}
\definecolor{diffdel}{rgb}{0.6,0.0,0.0}
\definecolor{diffcom}{gray}{0.4}
\definecolor{keywordblue}{rgb}{0.13,0.13,1}
\definecolor{stringgray}{rgb}{0.4,0.4,0.4}

\newenvironment{custombox}{\smallskip\begin{mdframed}[linewidth=1pt,innerleftmargin=5pt, innerrightmargin=5pt, innertopmargin=5pt, innerbottommargin=5pt, nobreak=true]}{\end{mdframed}\smallskip}

\usepackage{relsize}

\begin{document}

\title{Prompt Quality and Pull Request Outcomes: A Stage-Based Empirical Study of LLM-Assisted Development}

\author{\fnm{Richard} \sur{Sserunjogi}}\email{sserunjo@unlv.nevada.edu}

\author{\fnm{Daniel} \sur{Ogenrwot}}\email{ogenrwot@unlv.nevada.edu}

\author*{\fnm{John} \sur{Businge}}\email{john.businge@unlv.edu}

\affil{\orgdiv{Department of Computer Science}, \orgname{University of Nevada Las Vegas}, \orgaddress{\street{4505 S. Maryland Pkwy.}, \city{Las Vegas}, \postcode{89154}, \state{Nevada}, \country{United States of America}}}


\abstract{
Large language model (LLM)-powered tools such as ChatGPT are increasingly used in
collaborative software engineering workflows, yet little is known about
how prompt structure influences downstream pull request (PR) outcomes.
Prior studies primarily examine conversational helpfulness,
productivity, or coarse-grained adoption metrics, leaving the role of
prompt structure in collaborative integration behavior insufficiently
understood.
We analyze 265 manually validated developer--ChatGPT interactions
derived from self-admitted ChatGPT usage in open-source pull requests.
Building on prior research on developer-facing artifacts and prompt
engineering, we operationalize prompt structure using three
dimensions: \textit{Context}, \textit{Specificity}, and
\textit{Verification}. 
We first evaluate whether LLM-assisted annotation can reliably
reproduce human judgments of prompt structure, finding substantial
variation across prompt dimensions and workflow contexts.
\textit{Specificity} exhibits the most stable agreement with human
judgment, whereas \textit{Context} is systematically under-scored by
the LLM and \textit{Verification} remains difficult to assess
consistently, motivating a hybrid human--LLM annotation strategy.
Using this validated framework, we then examine how prompt structure
influences actionable code generation, code adoption, and integration
depth across AI-assisted PR workflows. 
\textit{Specificity} and \textit{Context} are most strongly associated with actionable code generation, \textit{Verification} emerges as the primary predictor of code adoption, and integration depth is most strongly associated with \textit{Context}, suggesting that deeper reuse depends on alignment with the surrounding implementation and project context.
Overall, our findings show that prompt characteristics exert distinct
stage-dependent effects across collaborative AI-assisted software
engineering workflows. The results suggest that prompt construction
functions as an important upstream factor in AI-assisted pull request
workflows, influencing not only code generation but also downstream
adoption and integration outcomes through contextual grounding,
task specificity, and evaluability cues.

}


\maketitle

\section{Introduction}

Large language model (LLM)-powered tools such as ChatGPT and GitHub Copilot are
increasingly shaping collaborative software engineering workflows by
supporting code generation, debugging, documentation, testing, and
pull request (PR) review
activities~\cite{
Russo:generativeAI:2024,
huang2024generative,
Ebert:GenAI:2023,
sauvola2024future}. Developers increasingly incorporate conversational
AI systems into day-to-day development tasks, often sharing these
interactions directly within GitHub issues and PR
discussions~\cite{tufano2026developers,
xiao2024generative,
hou2024large,Xiao:TSE:2026}. Prior work has examined the use of LLMs for code
generation, debugging, productivity improvement, and issue
resolution~\cite{peng2023impact,
siddiq2024using,
jin2024can,
Guo:2024,
Deng:2024,
Ehsani:EMSE:2026}. Recent studies further show that developers use
conversational AI for a broad range of software-engineering activities
extending beyond direct code
generation~\cite{tufano2026developers,Xiao:TSE:2026}.

Consider a common scenario in AI-assisted pull-request workflows: a developer asks an LLM to generate or revise code for a specific issue, but the usefulness of the response depends heavily on how the request is framed. Prompts with clear contextual grounding and constraints may produce actionable code that can be integrated with minimal revision, whereas vague or underspecified requests may yield incomplete, irrelevant, or unverifiable suggestions. During review and refinement, developers often revise prompts, clarify requirements, and provide additional context before deciding whether generated code should be adopted, adapted, or discarded. Prior work on ChatGPT-assisted pull requests shows that generated solutions are frequently refined, selectively reused, or substantially adapted rather than incorporated verbatim, underscoring the central role of developer evaluation and adaptation in AI-assisted workflows~\cite{ogenrwot2026patchtrack}. Prior work further shows that persistent interaction failures and declining confidence in LLM-generated outputs can lead developers to abandon AI assistance altogether~\cite{Tie:TOSEM:2026}. Consequently, the influence of LLMs on pull-request workflows is shaped not only by model capability, but also by the structure and quality of developer prompts.

Recent work characterizes prompts as first-class software artifacts within prompt-enabled systems and argues that prompt development requires systematic software-engineering methodologies beyond ad hoc trial-and-error practices~\cite{Chen:TOSEM:2026,Villamizar:PROFES:2025,Xiao:TSE:2026}. Complementary empirical work on prompt programming further shows that, when prompts are embedded in software systems and executed over variable inputs, developers must iteratively refine them, reason about model behavior, and address testing, debugging, and maintenance challenges that partly parallel traditional software development~\cite{Liang:FSE:2025}. Despite these advances, empirical understanding of how prompt characteristics influence collaborative AI-assisted development workflows remains limited. Existing studies primarily examine issue-resolution helpfulness, broad adoption or usage patterns, and generated-code quality~\cite{grewal2024analyzing,siddiq2024quality,Ehsani:EMSE:2026,Porta:EASE:2025}, leaving prompt-driven integration behavior within real-world pull request workflows insufficiently understood.

However, important challenges remain in AI-assisted software
development. LLM-generated outputs can contain hallucinations,
incomplete assumptions, and contextually inappropriate
implementations~\cite{tanzil2024chatgpt}. Multi-file and semantically
coordinated changes remain especially difficult due to limitations in
project-level reasoning and contextual
alignment~\cite{nashid2025characterizing}. These challenges highlight the need to better understand how prompt
characteristics influence downstream code-generation, adoption, and
integration outcomes in collaborative software-development workflows.

Prior work further suggests that AI-generated pull requests exhibit distinct communicative and integration characteristics compared to traditional human-authored contributions~\cite{Ogenrwot2026AICodingAgents,
Ogenrwot2026AgenticFlict}. Recent evidence also shows that AI-assisted pull requests involve complex review, adaptation, and integration processes that extend beyond direct code generation~\cite{ogenrwot2026patchtrack,Ogenrwot:2024:ASE}. Understanding how prompt structure influences code generation, adoption, and downstream integration is therefore essential to improving AI-assisted collaboration. Moreover, recent evidence suggests that advances in foundation models do not eliminate the need for prompt engineering entirely. While some sophisticated prompting strategies provide diminished benefits for advanced LLMs, execution-feedback-based and task-specific prompting techniques continue to improve performance on complex software-engineering tasks~\cite{Wang:TOSEM:2025}.

To address this gap, we develop and validate a framework for
operationalizing prompt structure in collaborative AI-assisted
software engineering. Building on prior research on bug-report
quality, pull request evaluation, testing practices, and prompt
engineering~\cite{bettenburg2008goodbugreport,
zimmermann2010goodbugreport,
gousios2016contributorperspective,
tsay2014githubfactors,
vasilescu2015ciquality,
zampetti2017externalreferences,
midolo2026promptguidelines},
we characterize prompt quality using three dimensions:
\textit{Context}, \textit{Specificity}, and
\textit{Verification} (CSV), which capture grounding, task
precision, and evaluability properties associated with effective
developer-facing artifacts. In this view, prompt structure represents
an additional layer of developer-facing communication that may
influence collaborative review and integration decisions alongside
traditional pull request factors.

Reliable operationalization of these constructs remains challenging,
however, because large-scale manual annotation of developer--LLM
interactions is costly and difficult to scale. Recent
software-engineering guidelines and broader annotation research
increasingly examine LLMs as annotators or evaluators, while
cautioning that LLM-generated labels can be sensitive to model
configuration, prompt design, task context, bias, and insufficient
human validation~\cite{baltes2025llmguidelines,wang:CHI:2024,
Takehi:SIGIR:2025,Xu:LAK26:2026}. We therefore first examine whether
LLM-based annotation can reliably reproduce human judgments of
prompt structure across the CSV dimensions and identify the
conditions under which selective annotation automation is
appropriate.

Building on this validated measurement framework, we model
AI-assisted pull request workflows as a multi-stage decision process.
Rather than treating AI assistance as a binary success-or-failure
outcome, we examine how prompt structure influences actionable code
generation, code adoption, and integration depth across successive
stages of real-world pull request workflows. Importantly, we do not
interpret adoption outcomes as direct indicators of intrinsic code
quality or model capability; rather, they reflect developer and
maintainer decision-making within collaborative pull request
workflows.

To study these dynamics systematically, we analyze a manually validated dataset of 265 pull requests that contain developer-shared links to ChatGPT conversations. These cases are drawn from self-admitted ChatGPT usage (SACU) in open-source repositories~\cite{ogenrwot2026patchtrack}. Unlike prior
studies that primarily examine conversational helpfulness,
developer adoption, or coarse-grained pull request outcomes, our
study focuses on how distinct prompt-quality dimensions influence
code generation, code adoption, and integration depth within
collaborative AI-assisted software development workflows. We further
evaluate whether LLM-assisted annotation can reliably reproduce
human judgments of prompt structure, motivating a hybrid annotation strategy that combines selective
automation with human oversight across different prompt dimensions
and workflow contexts.

Our analysis shows that different prompt-quality dimensions become
important at different stages of AI-assisted software development.
More concretely, \textit{Specificity} and \textit{Context} are most
strongly associated with actionable code generation,
\textit{Verification} is most strongly associated with code adoption,
and \textit{Context} again becomes the dominant factor for deeper
integration. These findings suggest that effective collaboration
depends not only on model capability or generated-code quality, but
also on how developers construct prompts. Prompt construction thus
emerges as a collaborative activity in which contextual grounding,
task specification, and evaluability cues shape downstream
development outcomes.

\vspace{5pt}
\noindent The paper makes the following contributions:
\begin{itemize}[leftmargin=*]
\setlength\itemsep{0em}

\item We develop a framework for operationalizing prompt structure in
AI-assisted software engineering using three dimensions: \textit{Context}, \textit{Specificity}, and
\textit{Verification} (CSV), enabling systematic analysis of
developer--LLM interactions in collaborative workflows.

\item We evaluate the reliability of LLM-assisted prompt annotation,
showing that annotation performance varies across prompt dimensions
and workflow contexts, and derive a dimension- and context-aware 
hybrid
human--LLM annotation strategy for scalable prompt-quality
assessment.

\item We introduce a stage-based workflow framework that relates
prompt structure to actionable code generation, code adoption, and
integration depth in AI-assisted pull request workflows.

\item We conduct a mixed-methods empirical study of 265
developer--ChatGPT interactions linked to open-source pull requests,
showing that different prompt-quality dimensions become important at
different stages of AI-assisted development workflows.

\item We release a public replication package to support
reproducibility and future research on AI-assisted collaborative
software engineering~\cite{sserunjogi_2026_20451326}.
\end{itemize}

\section{Operational Definitions and Scoring Rubric}
This section defines the key constructs used in our study and explains
how they are operationalized for analysis. We first introduce the
terminology used to describe ChatGPT-assisted interactions in pull
requests, then present the prompt-quality dimensions and their
associated scoring scheme, and finally describe the PR workflow
stages. Together, these constructs establish the foundation for the
analyses presented in subsequent sections.

\subsection{Terminology}

We adopt and restate key terminology from prior work on ChatGPT-assisted
pull request workflows~\citep{ogenrwot2026patchtrack}.
All definitions are provided here to ensure that this paper remains
self-contained.

\begin{itemize}[leftmargin=*, itemsep=1pt, topsep=2pt]

\item \textbf{Code Snippet:}
A small, standalone block of code intended for reuse, demonstrating a
specific task or programming concept. In this study, ``code snippet''
refers exclusively to discrete pieces of code suggested by ChatGPT,
excluding code examples embedded within descriptive explanations.

\item \textbf{Self-Admitted ChatGPT Usage (SACU):}
Instances where developers explicitly acknowledge, within a pull request
discussion, commit message, or issue thread, that they used ChatGPT to
generate, modify, or assist in writing code.

\item \textbf{Outcome Classes:}
Following  the work of Ogenrwot and Businge~\cite{ogenrwot2026patchtrack}, we classify ChatGPT-assisted pull requests into
four categories:

\begin{itemize}[leftmargin=*, itemsep=1pt, topsep=2pt]

\item \textbf{Patch Applied (PA):}
ChatGPT generates a code snippet, and the suggested code is incorporated
into the pull request, either directly or after some modification.
This outcome reflects cases where ChatGPT output contributes directly
to implementation, though often through adaptation rather than
verbatim reuse.

\item \textbf{Patch Not Applied (PN):}
ChatGPT generates a code snippet, but the suggested code is not
incorporated into the pull request, or is modified beyond recognizable
correspondence. Despite the lack of direct integration, these cases
often reflect meaningful developer engagement with the generated
solution, including adaptation to project constraints, exploration of
alternative implementations, or validation of design decisions.

\item \textbf{No Existing Patch (NE):}
The ChatGPT interaction provides textual guidance or explanation,
but does not produce a code snippet that can be evaluated as a patch.
These interactions nevertheless contribute to development by supporting
tasks such as debugging, conceptual clarification, documentation
refinement, and exploration of implementation strategies.

\item \textbf{Closed (CL):}
The pull request is closed without being merged, regardless of whether
ChatGPT-generated code was present. These cases may still involve code
generation or developer interaction with ChatGPT, but the pull request
does not result in integration into the codebase. Closure can arise from
a range of factors, including quality concerns, workflow constraints,
duplication, or misalignment with project requirements. \textit{In this
study, CL is treated as a pull request–level outcome and is analyzed
separately from code-focused analyses.}

\end{itemize}

\end{itemize}

\noindent These definitions establish the basis for characterizing prompt structure,
which we operationalize through three dimensions described next.
The outcome classes are later mapped to stages in the pull request
workflow, enabling analysis of how prompt structure influences code
generation, adoption, and integration.

\subsection{Prompt Quality Dimensions}

We ground our prompt-quality dimensions in prior software engineering
research on bug report quality, pull request evaluation, and
testing/CI
practices~\citep{bettenburg2008goodbugreport,
gousios2016contributorperspective,
kalliamvakou2015perilsgithub,
tsay2014githubfactors,
vasilescu2015ciquality,
zampetti2017externalreferences,
zimmermann2010goodbugreport}. Prior work highlights the importance
of contextual information, clear communication of developer intent,
and evidence supporting the correctness of proposed changes when
developers create, evaluate, and integrate software artifacts.
Drawing on these recurring themes, we operationalize prompt quality
using three dimensions: Context, Specificity, and Verification.

We score each dimension on a common three-point ordinal scale (0--2), where 0 indicates absence of the relevant cue, 1 indicates partial or implicit presence, and 2 indicates explicit and well-developed presence. This scale reflects the recurring distinction in prior software-engineering work between missing, incomplete, and actionable information in developer-facing artifacts, including bug reports, pull requests, tests, CI signals, and external references~\citep{bettenburg2008goodbugreport,zimmermann2010goodbugreport,tsay2014githubfactors,vasilescu2015ciquality,zampetti2017externalreferences}. Using the same scale across Context, Specificity, and Verification provides a consistent representation of prompt quality while limiting annotation ambiguity.

\paragraph{Context (C)}
Prior research on bug reports and pull request evaluation highlights the importance of concrete technical information and supporting evidence when diagnosing and assessing software changes~\citep{bettenburg2008goodbugreport,zimmermann2010goodbugreport,tsay2014githubfactors}. We define \textbf{Context} as the extent to which a prompt is grounded in concrete technical artifacts. High-context prompts identify the locus of change and provide supporting evidence such as code snippets, logs, or error messages. We assign $C=2$ when both are present, $C=1$ when only one is provided, and $C=0$ when neither is present.

\paragraph{Specificity (S)}
Studies of pull-based development and prompt engineering emphasize the importance of clearly communicating intended changes, constraints, and expectations~\citep{gousios2016contributorperspective,midolo2026promptguidelines}. We define \textbf{Specificity} as the clarity and precision with which a prompt articulates the intended task. High-specificity prompts define the goal, scope, and relevant constraints, such as API requirements, output formats, or example behaviors. We assign $S=2$ when all three are specified, $S=1$ when scope or constraints are incomplete, and $S=0$ when the prompt is vague or underspecified.

\paragraph{Verification (V)}
Evidence such as tests, CI outcomes, examples, and external references is commonly used to assess whether software changes satisfy expected behavior~\citep{vasilescu2015ciquality,zampetti2017externalreferences,midolo2026promptguidelines}. We define \textbf{Verification} as the presence of signals that allow developers to assess whether generated outputs are valid. Such signals may include input--output examples, rules, tests, or expected behaviors. We assign $V=2$ when the prompt includes a concrete, checkable evaluation criterion, $V=1$ when expected behavior is implied but not explicitly defined, and $V=0$ when no evaluation signal is present.

We also considered \textbf{Prompt Efficiency} as a potential dimension, capturing the number of interaction turns required to obtain a satisfactory LLM-generated solution. However, we excluded this dimension due to construct validity concerns. ChatGPT conversations shared in pull request discussions represent only the subset of interactions developers choose to disclose and may omit exploratory prompting performed offline. Developers may therefore share only the final successful interaction, making observed turn counts an unreliable proxy for actual effort. Preliminary analysis further revealed no stable relationship with adoption outcomes. We therefore focus on structural prompt properties (Context, Specificity, and Verification) that can be measured consistently from shared artifacts.

\subsection{PR Workflow Stages}

To analyze how prompt structure influences pull request outcomes, we
conceptualize the development process as a sequence of stages reflecting
how ChatGPT-generated content is used in practice. These stages are
operationalized using the PR outcome classes defined earlier.

\begin{itemize}[leftmargin=*]

\item \textbf{Code Generation:}
Whether a prompt leads ChatGPT to produce a code snippet that can be
considered for reuse. This stage distinguishes between interactions
where no code is generated (\textit{NE}) and those where code is produced
(\textit{PA} $\&$ \textit{PN}).

\item \textbf{Adoption:}
Whether the generated code is incorporated into the pull request,
either directly or after modification. This stage distinguishes between
cases where generated code is adopted (\textit{PA}) and those where it is
not (\textit{PN}).

\item \textbf{Integration depth:}
The extent to which generated code contributes to the final
implementation. This stage focuses on \textit{PA} cases and captures
variation in the degree of incorporation rather than binary adoption.

\end{itemize}

In addition to these code-level stages, we consider \textbf{pull request
lifecycle outcomes}, which capture whether a pull request is ultimately
merged or closed (\textit{CL}), as well as the time required to reach
resolution.

\subsection{Concrete Example}

To illustrate the constructs used in this study, we present a
real-world example of a ChatGPT-assisted pull request interaction. In
this case, a developer sought to eliminate duplication in a TypeScript
discriminated union by deriving the union from a mapping type. The
prompt includes the original type definitions and asks whether a more
concise formulation is possible. The full interaction is available
online\footnote{\url{https://chat.openai.com/share/f09f38e5-f541-4f98-9483-e183f5650398}}.
ChatGPT responded by introducing a mapped type and indexed-access
pattern that removed the need for manual enumeration. Examination of
the corresponding pull
request\footnote{\url{https://github.com/nbd-wtf/nostr-tools/pull/241}}
shows that the solution was largely incorporated into the repository
with only minor adaptations. Using \texttt{PatchTrack}, an analytical
instrument for classifying patch outcomes and usage patterns under AI
assistance~\cite{ogenrwot2026patchtrack}, we estimate that
approximately 60\% of the ChatGPT-generated code was retained in the
final version.

From the perspective of our prompt-quality dimensions, this
interaction exhibits strong contextual grounding because the prompt
includes concrete code and clearly specifies the transformation goal;
partial specificity because the desired outcome is stated without
fully constraining the solution; and implicit verification because
correctness is inferred from type equivalence rather than explicit
tests or formal specifications. This example illustrates the type of
prompt--response interaction analyzed in this study and motivates our
investigation of how prompt structure influences code generation,
adoption, and integration outcomes.
\begin{figure*}
\centering
\includegraphics[width=0.8\linewidth]{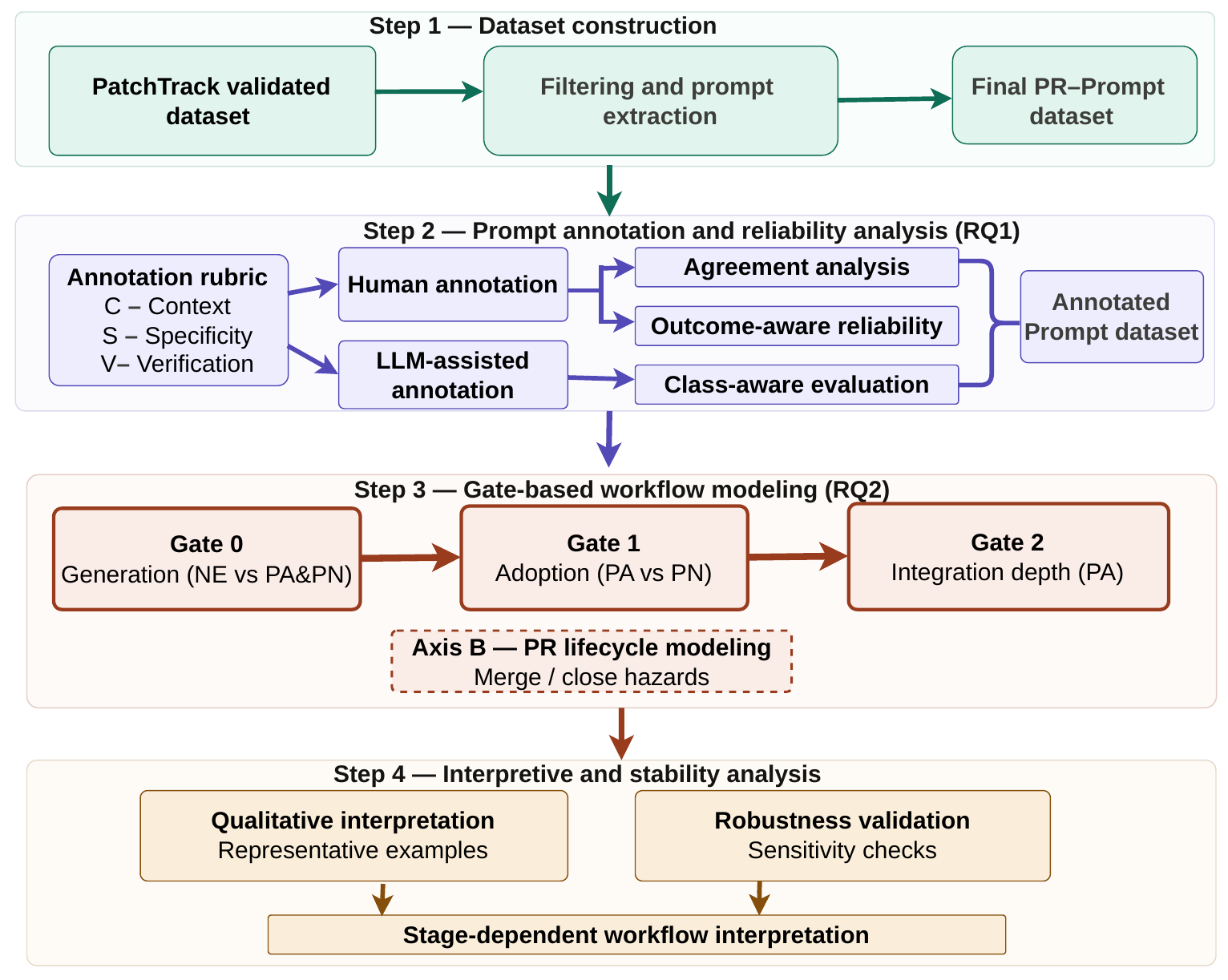}
\caption{Overview of the study design.}
\label{fig:method}
\vspace{-10pt}
\end{figure*}

\section{Study Design}

This section describes the overall research methodology, including
dataset construction, prompt-quality measurement, and the
quantitative and qualitative analyses used to investigate the
relationship between prompt structure and pull request outcomes.
The study proceeds in four stages. First, we construct the
PR--prompt dataset and prepare it for analysis. Second, we annotate
prompts using the Context, Specificity, and Verification (CSV)
framework and evaluate the reliability of LLM-assisted annotation.
Third, we model AI-assisted pull request workflows using a
gate-based framework that examines code generation, adoption,
integration depth, and pull request lifecycle outcomes. Finally, we
conduct qualitative and robustness analyses to interpret and validate
the observed stage-dependent relationships. Figure~\ref{fig:method}
provides an overview of the study design.

\subsection{Analytical Scope and Inclusion Criteria}
\label{sec:analytical-scope}

Our analysis distinguishes between prompt-level interactions and
pull request level outcomes to isolate the effects of prompt
structure from broader project and workflow influences. We focus on
stages corresponding to code generation, adoption, and integration,
while treating overall pull request lifecycle outcomes separately.

The outcome classes \textit{PA}, \textit{PN}, and \textit{NE} map
directly to these code-level stages and therefore form the basis of
our gate-level analysis. Specifically, \textit{NE} captures cases
where no code is generated, \textit{PN} captures cases where
generated code is not adopted, and \textit{PA} represents cases
where generated code is incorporated into the pull request.

In contrast, \textit{Closed (CL)} pull requests may still involve
high-quality prompts and ChatGPT-generated code, but closure often
reflects factors external to prompt structure, such as project
policies, workflow constraints, duplicate effort, or reviewer
decisions. Including CL cases in gate-level models would therefore
conflate prompt-level effects with project-level influences.
Accordingly, we restrict gate-level analyses to \textit{PA},
\textit{PN}, and \textit{NE} cases, while analyzing CL cases
separately as part of pull request lifecycle outcomes (Axis~B).

\subsection{Research Questions}

This study investigates how the structural properties of developer
prompts influence progression through AI-assisted pull request
workflows. We first evaluate the reliability of prompt-structure
measurement, and then analyze how validated prompt dimensions relate
to code generation, adoption, integration depth, and pull request
lifecycle outcomes.

\begin{itemize}[leftmargin=*, itemsep=4pt]

\item \textbf{RQ1:} \textit{How reliably can large language models reproduce human judgments of prompt structure?}

\item[] Reliable operationalization of prompt-quality constructs is essential
for large-scale empirical studies of developer--LLM interactions.
This question evaluates whether LLM-based annotation can reproduce
human judgments of Context, Specificity, and Verification, and
whether annotation reliability varies across workflow contexts in ways
that affect the design of scalable human--LLM annotation strategies.

\begin{itemize}[leftmargin=*, itemsep=2pt]

\item \textbf{RQ1a (Overall Agreement):}
\textit{To what extent do LLM-generated annotations align with human
consensus across prompt dimensions, and do systematic biases emerge?}

\item \textbf{RQ1b (Outcome-Conditioned Agreement):}
\textit{Does annotation agreement vary across pull request outcome classes (PA, PN, NE, CL)?}

\item[] Variations in agreement may indicate that certain workflow
contexts are inherently more difficult to assess.

\item \textbf{RQ1c (Hybrid Annotation Policy):}
\textit{Which prompt dimensions and workflow contexts are suitable
for LLM-assisted annotation without compromising construct validity?}

\item[] This question informs the design of a class-aware hybrid
annotation strategy balancing scalability with human oversight.

\end{itemize}

\item \textbf{RQ2:} \textit{How does prompt structural quality influence progression through the pull request integration pipeline?}

\item[] Using validated annotations, we examine how prompt structure
relates to code generation, adoption, integration depth, and broader
pull request lifecycle dynamics.

\begin{itemize}[leftmargin=*, itemsep=2pt]

\item \textbf{RQ2a (Code Generation):}
\textit{Does prompt structure influence whether ChatGPT produces actionable code (NE vs.\ non-NE)?}

\item \textbf{RQ2b (Adoption):}
\textit{Does prompt structure influence whether generated code is adopted in the pull request (PA vs.\ PN)?}

\item[] This stage reflects whether generated code is considered
sufficiently useful and trustworthy for inclusion.

\item \textbf{RQ2c (Extent of Integration Depth:}
\textit{Among adopted cases, does prompt structure relate to the fraction of ChatGPT-generated code integrated?}

\item[] This stage distinguishes superficial reuse from deeper
integration.

\item \textbf{RQ2d (Lifecycle Outcomes):}
\textit{Does prompt structure relate to pull request outcomes (merged vs.\ closed) and time-to-resolution?}

\end{itemize}

\end{itemize}

\subsection{Dataset Construction and Prompt Extraction}
\label{sec:data-construction}

To address our research questions, we construct a dataset of
developer--ChatGPT interactions grounded in real-world pull request
(PR) workflows. Our approach builds on the growing use of
self-admitted GenAI usage artifacts as an empirical data source for
studying developer interactions with conversational AI systems in
open-source software development~\cite{Xiao:TSE:2026}. We build on
the manually validated \texttt{PatchTrack}
corpus~\cite{ogenrwot2026patchtrack}, which contains pull requests with developer-shared links to ChatGPT conversations classified according to how generated code was
used during development. By refining this validated corpus for
prompt-level analysis, we obtain a traceable dataset linking
prompts, generated code, and pull request outcomes, enabling
systematic analysis of how prompt structure relates to code
generation, adoption, integration depth, and broader workflow
outcomes.

\paragraph{\textbf{Step 1:} Source Dataset and Outcome Classification}

The original \texttt{PatchTrack} validated dataset contains 338 pull
requests with developer-shared ChatGPT conversations. In the
\texttt{PatchTrack} study, each PR was classified according to how
ChatGPT-generated code was used within the development workflow using
token-based similarity analysis between generated snippets and pull
request diffs~\cite{pareco:2022}. Based on this analysis, each PR was assigned one of
four outcome classes: Patch Applied (PA), Patch Not Applied (PN), No
Code Patch / Guidance Only (NE), and Closed Unmerged (CL). To ensure
reliability, all classifications were manually validated through
inspection of ChatGPT conversations, pull request discussions, and
code diffs, resulting in a high-confidence subset of 283 validated
cases (PA-89, PN-63, NE-84, and CL-47).

\paragraph{\textbf{Step 2:} Selection of Validated Subset}

To ensure consistency and avoid propagating classification errors, we
restrict our analysis to this manually validated subset. We treat the
\texttt{PatchTrack} labels as ground truth and do not re-derive or modify them in
this study. The validated dataset provides a balanced representation of
adoption, non-adoption, guidance-only interactions, and closed pull
requests, enabling robust analysis across outcome types.

\paragraph{\textbf{Step 3:} Prompt Availability Filtering}

Because our analysis focuses on prompt quality, we further filter the
validated dataset based on the availability of developer prompts. In
some cases, the original ChatGPT shared links were no longer accessible
(e.g., deleted or inactive), preventing reliable extraction of prompt
content. These cases were removed to ensure that all included instances
contain complete and analyzable prompt data.
After filtering, the final dataset consists of 265 PR-linked cases.
The reduction is primarily due to missing prompt data in a subset of
Patch Not Applied (PN) cases,  with additional exclusions in the No Existing Patch (NE) and Closed (CL) categories (PA-89, PN-53, NE-80, and CL-43). 

\paragraph{\textbf{Step 4:} Prompt Extraction and Data Representation}

For each remaining case, we extract the developer prompt text from the
ChatGPT conversation, including any refinements present within the
interaction. Each case is represented as a tuple consisting of the
prompt text, the corresponding PR outcome class, and traceability links
to both the ChatGPT conversation and the GitHub pull request.
Each instance is assigned a unique \textit{Case ID}, which serves as the
primary key throughout all stages of analysis, including annotation,
validation, and modeling.

\paragraph{\textbf{Step 5:} Data Integrity and Reproducibility}

All preprocessing steps, including dataset filtering, prompt extraction,
and case mapping, are documented in the replication package to ensure
full transparency and reproducibility. Our
procedures follow established guidelines for empirical software
engineering studies involving LLMs, emphasizing artifact preservation,
traceability, and reproducible data preparation~\cite{baltes2025llmguidelines}.

\subsection{Prompt Evaluation and Validation Methodology}
\label{sec:prompt-eval}

To address \textbf{RQ1}, we design a structured methodology to
(i) operationalize prompt quality, (ii) evaluate the reliability of
LLM-based annotation, and (iii) derive a validated annotation strategy
for large-scale analysis. Our approach integrates rubric-based scoring,
LLM-assisted annotation, and human validation to ensure both scalability
and measurement validity.

\subsubsection{Prompt Quality Annotation}

We operationalize prompt quality using a structured rubric capturing
three dimensions: \textit{Context}, \textit{Specificity}, and
\textit{Verification} (\textit{CSV}). Each dimension is scored on a three-point
ordinal scale (0--2), reflecting the extent to which a prompt is
grounded in technical artifacts, clearly specifies the intended task,
and provides verifiable correctness criteria. Following recommendations
for LLM-assisted empirical studies~\cite{baltes2025llmguidelines}, we
adopt a hybrid human--LLM annotation workflow that combines human
validation with automated annotation to balance scalability and
construct validity.

A stratified 30\% subset of prompts from each outcome class (CL: $n=13$, PN: $n=16$, PA: $n=27$, and NE: $n=24$) was
independently annotated by two PhD-level researchers using the CSV
rubric. To minimize potential bias, annotators were blinded to all
LLM-generated labels during both annotation and adjudication. Because
prompt-quality labels are subsequently used in all downstream analyses,
establishing annotation reliability is a necessary prerequisite for
construct validity. Prior to any discussion or reconciliation,
inter-rater agreement was assessed using weighted Cohen's Kappa
($\kappa_w$), which is appropriate for ordinal annotation tasks.
Following reliability assessment, disagreements were resolved through
discussion, and unresolved cases were adjudicated by a senior
researcher (third co-author). The resulting consensus labels
constituted the gold-standard reference used in all subsequent
Human--LLM agreement analyses.

To support transparency and reproducibility, the replication package
contains the annotation rubric, annotation instructions, independent
annotator labels, disagreement records, adjudication decisions,
consensus labels, annotation prompts, evaluation logs, and
configuration details~\cite{sserunjogi_2026_20451326}. These materials are
provided in accordance with recommended reporting practices for
LLM-assisted empirical studies~\cite{baltes2025llmguidelines}.

\subsubsection{LLM Annotation Procedure}

LLM-based annotation was performed using a structured prompt (LLM-v1) that operationalizes the Context--Specificity--Verification (CSV) rubric. For each interaction, LLM-v1 receives the developer-authored prompt extracted from the shared ChatGPT conversation together with the rubric definitions and scoring criteria. The model assigns an ordinal score (0--2) for each dimension and provides a brief justification. Human annotators were blinded to all LLM-generated labels during annotation and adjudication to prevent automated scores from influencing construction of the gold-standard dataset.

Annotations were generated using ChatGPT under its default configuration, including default decoding parameters. Annotation experiments were conducted between 14~November~2025 and 21~January~2026. During this period, ChatGPT 5.1 generated annotations for 222 interactions and ChatGPT 5.2 generated annotations for 43 interactions. To assess the impact of the model transition, we compared annotations from both versions against the human-adjudicated gold standard. No systematic differences in scoring behavior or rubric interpretation were observed, suggesting that the version change did not materially affect the reported results.

\subsubsection{Human--LLM Agreement Evaluation}
\label{sec:agreement}
After establishing the human gold standard, we evaluate whether
LLM-generated annotations can reliably reproduce human judgments of
prompt quality. LLM-generated scores are compared against the consensus
human labels on the gold-standard subset.

Because the CSV dimensions are measured on an ordinal three-point scale
(0--2), we use complementary metrics that capture both agreement and
error characteristics. Specifically, we compute quadratic weighted
Cohen's $\kappa_w$ to measure chance-adjusted ordinal agreement while
assigning partial credit to adjacent-category disagreements
~\citep{cohen1968weighted,fleiss1981statistical}. We additionally
report mean absolute error (MAE) to quantify the average magnitude of
scoring deviations~\citep{willmott2005advantages} and directional bias
to identify systematic over- or under-scoring relative to the human
reference labels.

To assess robustness across pull request contexts, we additionally
compute class-specific agreement using quadratic weighted $\kappa_w$
for each outcome class (PA, PN, NE, CL). Following common practice,
commonly cited Kappa interpretation
categories~\citep{landis1977measurement} are used only as descriptive
reference points rather than strict decision thresholds. Annotation
reliability is instead assessed using the combined evidence provided
by agreement statistics, error magnitude, directional bias, and
qualitative error analysis.

Beyond aggregate agreement metrics, we examine representative
Human--LLM disagreement cases to better understand the sources of
annotation variation. Specifically, we review prompts exhibiting
substantial disagreement to assess whether misalignment reflects
boundary ambiguity between adjacent categories, differences in rubric
interpretation, or systematic scoring tendencies. This diagnostic
analysis contextualizes the quantitative results and is motivated by
recent work showing that aggregate agreement metrics alone can obscure
important differences between annotation ambiguity, systematic model
bias, and context-dependent labeling failures in LLM-assisted
annotation workflows~\citep{Xu:LAK26:2026,wang:CHI:2024,Takehi:SIGIR:2025}.

\subsubsection{Annotation Policy Derivation}

The agreement analysis described in Section~\ref{sec:agreement} was
used to determine which prompt dimensions and outcome classes could be
reliably annotated using LLM assistance and which required continued
human evaluation. Rather than assuming uniform reliability across
workflow contexts, we adopt a selective verification strategy in which
LLM-generated labels are accepted only when agreement with the human
gold standard is sufficient, while dimensions exhibiting persistent
disagreement remain human-scored. This design follows recent
recommendations for hybrid human--LLM annotation workflows that
combine automated labeling with targeted human validation for
unreliable cases~\cite{wang:CHI:2024,Takehi:SIGIR:2025}.

The policy was derived using the combined evidence from agreement
strength, error magnitude, directional bias, and qualitative
disagreement analysis rather than any single metric in isolation.
Consequently, dimensions and workflow contexts exhibiting reduced
reliability were retained as human-scored, yielding a conservative,
class-aware, and dimension-specific annotation strategy that balances
scalability, reliability, and construct validity. The resulting policy
was subsequently applied to construct the final prompt-quality dataset
used in RQ2, with annotations assigned through either LLM-assisted or
human evaluation according to the reliability criteria established
above.

\subsection{Modeling Prompt Effectiveness Across Development Stages}
\label{sec:modeling}

To address \textbf{RQ2}, we adopt a structured modeling framework to
evaluate how prompt structure influences AI-assisted pull request
outcomes across multiple stages of the development workflow. Prompt
quality is operationalized using a Prompt Quality Score (PQS),
defined over three dimensions: \textit{Context (C)},
\textit{Specificity (S)}, and \textit{Verification (V)}. Each
dimension is scored on a three-point ordinal scale (0--2) and
aggregated as:

\begin{equation}
\text{PQS} = C + S + V \in [0,6]
\label{eq:pqs}
\end{equation}

While PQS provides a summary measure, all primary models are
specified using the individual dimensions (C, S, V) to preserve
interpretability and isolate their respective effects.

We model prompt effectiveness as a sequence of stages reflecting how
LLM outputs are generated and incorporated into pull requests:
\textbf{Gate~0} captures whether a prompt produces code,
\textbf{Gate~1} whether generated code is adopted, and
\textbf{Gate~2} the extent to which generated code is integrated.
This decomposition separates code generation, adoption, and reuse
rather than treating AI assistance as a single aggregated outcome.
We additionally consider \textbf{Axis~B} (Pull Request Lifecycle
Outcomes), which captures pull request resolution status and
time-to-resolution separately from the gate-level workflow analysis.

Regression-based models are aligned with the structure of each stage. Logistic regression is used for binary outcomes (Gates~0 and~1), while models appropriate for bounded proportional data are used for Gate~2. All models include the prompt dimensions (C, S, V). To account for pull request complexity, we include Log(PR Size) as a control variable. PR Size is measured as total code churn, computed as the sum of lines added and deleted in a pull request, and is log-transformed to reduce skewness~\cite{businge:2018icsme,businge:saner:2022,Businge:EMSE:2023}

\subsubsection{Model Diagnostics}

To evaluate the suitability of the statistical models, we conducted a
series of diagnostic checks. Multicollinearity among predictors was
assessed using variance inflation factors (VIF), following established
regression diagnostics practices~\citep{kutner2005applied,
james2013introduction}. Logistic regression models were examined for
complete or quasi-separation, which can lead to unstable coefficient
estimates~\citep{albert1984existence}. For time-to-event analyses
associated with Axis~B, proportional hazards assumptions were evaluated
using Schoenfeld residuals~\citep{schoenfeld1982residuals,
grambsch1994proportional}. We additionally conducted robustness and
sensitivity analyses using alternative model specifications and reduced
predictor sets to assess the stability of the findings.

\subsubsection{Robustness and Sensitivity Analysis}
\label{sec:robustness}

To assess the stability of the gate-level findings, we conduct a
series of robustness and sensitivity analyses targeting repository
heterogeneity, skewed pull request distributions, dominant technical
domains, and alternative outcome operationalizations. We evaluate
alternative model specifications, including repository-clustered
standard errors and mixed-effects models with repository-level random
effects, to account for potential within-repository dependence.

We further repeat the analyses after excluding dominant repositories,
extreme pull request sizes, and the most frequent programming
languages to assess potential outlier and language-specific effects.
Additional analyses examine alternative operationalizations of
integration depth in Gate~2, compare aggregate Prompt Quality Score
(PQS) models against individual CSV dimensions, and incorporate prompt-length controls and
hold-out stability checks where applicable. Across these analyses,
our focus is on the qualitative stability of stage-dependent
relationships rather than exact coefficient replication.

\subsubsection{Directed Qualitative Content Analysis}
\label{sec:method-qual}
To complement the quantitative modeling, we conduct a directed
qualitative content analysis~\cite{Hsieh2005} to interpret how prompt
structure relates to different stages of the pull request workflow.
The analysis uses the Context, Specificity, and Verification rubric to
guide examination of prompt rationale descriptions and associated
developer--LLM interactions. Rather than seeking an exhaustive thematic
characterization of developer--LLM interactions, we adopt a targeted
interpretive approach to explain and contextualize the statistical
relationships observed in the quantitative analysis.

For each workflow stage, the reviewers examined a small set of representative and contrasting cases (approximately ten per stage) spanning high-, medium-, and low-scoring prompt configurations, multiple outcome classes, technical domains, and workflow outcomes. Sampling intentionally included cases that both aligned with and diverged from the dominant quantitative trends to reduce reliance on confirmatory examples and surface alternative explanations where applicable. Through cross-case comparison, the reviewers identified recurring patterns in how prompts formulate implementation tasks, express constraints, provide contextual grounding, and support evaluation-related reasoning. Initial interpretations were refined by revisiting the original prompts and conversations, and disagreements were resolved through iterative discussion and consensus refinement.

The qualitative analysis was conducted across all workflow stages, and
representative examples were selected only after recurring patterns had
been identified through cross-case review. Accordingly, the examples
are intended to illustrate and contextualize the quantitative findings
rather than serve as standalone evidence or support statistical
generalization. Additional examples exhibiting similar characteristics
are included in the replication package to support transparency and
cross-case comparison.

\section{Results}
\label{sec:results}

This section presents the key findings of our study. We first evaluate
the reliability of LLM-based prompt annotation (RQ1) and then examine
how validated prompt dimensions relate to progression through the
pull request workflow (RQ2).

\subsection{RQ1: Reliability of LLM-Based Prompt Annotation}
\label{sec:rq1_results}

We assess the extent to which LLM-based annotation reproduces human
judgments of prompt structural quality and whether such annotations
can support downstream modeling.

\paragraph{Gold-standard annotation reliability.}
As described in Section~\ref{sec:prompt-eval}, the gold-standard dataset was established through independent human annotation followed by adjudication. Prior to reconciliation, inter-rater agreement was assessed using weighted Cohen's Kappa ($\kappa_w$) to evaluate the reliability of the CSV rubric across the validation subset.

Agreement varied substantially across dimensions and outcome classes.
Context agreement ranged from $\kappa_w=-0.059$ (CL) to
$\kappa_w=0.563$ (NE), Specificity from $\kappa_w=-0.573$ (CL) to
$\kappa_w=0.440$ (NE), and Verification from $\kappa_w=-0.043$ (NE)
to $\kappa_w=0.467$ (PA). Raw agreement ranged from 23.1\% to 91.7\%
across the evaluated combinations. The weakest agreement was observed
primarily in CL and, to a lesser extent, NE, where small validation
subsets and highly imbalanced score distributions contributed to
unstable $\kappa_w$ estimates.

These findings indicate that annotation reliability varies substantially across prompt dimensions and workflow contexts, underscoring the challenges of prompt-quality assessment and motivating the class-aware evaluation strategy developed in RQ1c. Following reliability assessment, all disagreements were resolved through discussion and adjudication, and the resulting consensus labels served as the gold-standard reference for subsequent Human--LLM agreement analyses.

\subsubsection{RQ1a: Overall Human--LLM Agreement}

LLM-based annotation exhibits moderate agreement with human judgment
for both \textit{Context} and \textit{Specificity}. However,
\textit{Context} exhibits a consistent under-scoring tendency relative
to human annotations, whereas \textit{Verification} achieves only fair
agreement, indicating greater difficulty in operationalizing
verification-related signals.
We evaluate agreement between annotations produced by LLM-v1 and
blinded human consensus annotations across the human-validation subset
($N=80$).
Context achieves moderate agreement ($\kappa_w = 0.526$), but with a
relatively high mean absolute error (MAE = 0.475) and substantial
negative bias ($-0.450$), indicating that LLM-v1 systematically
assigns lower Context scores than the human reference labels. 
Specificity also exhibits moderate agreement
($\kappa_w = 0.431$), with lower error (MAE = 0.325) and a smaller
negative bias ($-0.200$). Verification shows only fair agreement
($\kappa_w = 0.316$), with comparable error (MAE = 0.338) and minimal
directional bias ($-0.037$). Overall, these results suggest that LLM-based annotation more reliably
captures observable prompt characteristics than implicit correctness
and verification cues.

\paragraph{Explaining Agreement Patterns.}
The stronger agreement observed for \textit{Context} and
\textit{Specificity} is consistent with their reliance on explicit
structural features such as code artifacts, system details, and task
constraints that are directly observable in the prompt text. In
contrast, \textit{Verification} often depends on implicit correctness
expectations and evaluative reasoning that may be underspecified or
distributed across the broader conversational context, making
consistent interpretation more difficult. The negative bias observed
for \textit{Context} further suggests that LLM-v1 requires more
explicit contextual evidence than human annotators before assigning
higher Context scores.

\subsubsection{RQ1b: Outcome-Conditioned Agreement Patterns}
Agreement between LLM and human annotations varies across outcome classes, particularly for Verification. To assess robustness across workflow contexts, we compute quadratic weighted $\kappa_w$ separately for each outcome class and dimension
(Table~\ref{tab:per_class_kappa}).
\begin{table}[h]
\centering
\begin{tabular}{lcccc}
\toprule
Class & $N$ & Context $\kappa_w$ & Specificity $\kappa_w$ & Verification $\kappa_w$ \\
\midrule
PA & 27 & 0.427 & 0.053 & 0.073 \\
PN & 16 & 0.308 & 0.282 & 0.667 \\
CL & 13 & 0.544 & 0.581 & -0.020 \\
NE & 24 & 0.621 & 0.438 & -0.075 \\
\bottomrule
\end{tabular}
\vspace{-4pt}
\caption{Per-class quadratic weighted $\kappa_w$ values.}
\label{tab:per_class_kappa}

\vspace{-15pt}
\end{table}

Agreement varies markedly across outcome classes. Verification shows very low agreement in PA ($\kappa_w = 0.073$) and negative agreement in CL and NE, indicating substantial human--LLM disagreement on verification-related cues. Specificity also shows very low agreement in PA ($\kappa_w = 0.053$), but reaches moderate levels in CL and NE. Context is more stable, with moderate-to-substantial agreement in CL ($\kappa_w = 0.544$) and NE ($\kappa_w = 0.621$), though weaker in PN. Overall, annotation reliability is outcome-dependent, with some dimensions harder to score consistently in particular workflow contexts.

\paragraph{Explaining Outcome-Conditioned Agreement.}
The observed variation across outcome classes likely reflects differences in how information and correctness expectations are expressed within prompts. During annotation and adjudication, NE and CL prompts often contained localized structural cues, making Context and Specificity easier to identify consistently. In contrast, PA prompts frequently involved iterative refinement, partial implementations, or loosely specified requirements, causing relevant signals to be more distributed and difficult to interpret.

Verification agreement appears particularly sensitive to the clarity of correctness cues. In PA cases, verification signals are often implicit and must be inferred from the interaction, contributing to very low agreement. By contrast, PN cases frequently contain explicit evaluation or rejection reasoning, consistent with the substantially higher Verification agreement ($\kappa_w = 0.667$).

\subsubsection{RQ1c: Deriving an Empirically Grounded Evaluation Policy}

The RQ1a and RQ1b results indicate that annotation reliability varies
substantially across prompt dimensions and outcome classes. As a
result, no single automation strategy achieves consistent agreement
across all workflow contexts, motivating a class-aware evaluation
policy. 
The recommended annotation policy was derived primarily from agreement strength ($\kappa_w$) and evidence of systematic over- or under-scoring identified through the bias analysis, with MAE used as a supplementary indicator of scoring behavior.
Following widely used interpretations of Cohen's $\kappa_w$, values of $\kappa_w \geq 0.40$ were used as a reference benchmark for identifying dimensions that achieved at least moderate agreement~\cite{landis1977measurement,gonzalez2023rel,diaz2023applying}.
We therefore derived the dimension- and class-aware evaluation policy shown in Table~\ref{tab:recommended_policy} using the combined evidence from agreement strength, error magnitude, directional bias, and qualitative examination of disagreement patterns. The class-specific agreement estimates are based on stratified subsets of the dataset and should therefore be interpreted as indicative of relative annotation reliability across workflow contexts rather than as precise population parameters. Accordingly, the policy should be viewed as a conservative recommendation informed by multiple sources of evidence rather than a strict threshold-based decision rule.

\begin{table}[h]
\centering
\begin{tabular}{lcccc}
\toprule
Metric & PA & PN & CL & NE \\
\midrule
Context & Human & Human & Human & Human \\
Specificity & Human & Human & LLM & LLM \\
Verification & Human & LLM & Human & Human \\
\bottomrule
\end{tabular}

\vspace{-4pt}
\caption{Recommended human--LLM annotation policy derived from agreement and bias analyses.}
\label{tab:recommended_policy}

\vspace{-20pt}
\end{table}

Context is retained as human-scored across all outcome classes because
LLM-v1 consistently assigns lower Context scores than human
annotators despite moderate agreement levels. In contrast,
LLM-assisted scoring is used selectively for Specificity (CL and NE)
and Verification (PN) in workflow contexts where agreement is
sufficiently reliable, while all remaining dimension--outcome
combinations remain human-scored.

\paragraph{Interpreting the Evaluation Policy.}
The resulting policy highlights that annotation reliability depends on
both prompt dimension and workflow context, motivating selective human
oversight rather than uniform automation.

\begin{custombox}
{\faLightbulbO \hspace{0.01in} \textbf{Summary of RQ1 Results:}}
LLM-based annotation achieves moderate agreement with human judgments
for some prompt dimensions, but reliability varies substantially
across prompt dimensions and workflow contexts.
\textit{Context} and \textit{Specificity} achieve moderate agreement
overall, although \textit{Context} consistently receives lower scores
from the LLM and \textit{Specificity} exhibits substantial variation
across outcome classes. \textit{Verification} remains difficult to
assess when correctness expectations are implicit. Agreement also
varies across outcome classes, indicating that annotation reliability
depends not only on the construct being measured but also on the
surrounding workflow context.
These findings suggest that uniform annotation automation is
inappropriate. Instead, reliable operationalization of prompt
structure requires a class-aware hybrid human--LLM strategy in which
LLM-assisted scoring is selectively applied only in
dimension--outcome contexts exhibiting acceptable agreement and
limited systematic bias, while human oversight is retained for
contexts that remain difficult to annotate consistently.
\end{custombox}

\subsection{RQ2: Prompt Structure and Progression Through the Pull Request Workflow}
\label{sec:rq2_results}

Building on the validated prompt-quality dataset produced through the class-aware hybrid annotation strategy developed in RQ1, we analyze how prompt structure influences progression through the pull request workflow.
We conceptualize this process as
a sequence of stages corresponding to code generation, adoption,
and integration, followed by broader pull request lifecycle
outcomes. This stage-based framework allows us to examine how the
role of prompt structure evolves across the development workflow.

We combine quantitative modeling with targeted qualitative analysis
to interpret the observed effects. Quantitative results identify
statistical relationships between prompt dimensions and workflow
outcomes, while illustrative examples are used to explain the
mechanisms underlying these patterns. Together, these analyses
provide insight into how prompt structure shapes LLM-assisted
development in practice.

\subsubsection{Descriptive Statistics}
We summarize the distributions of Context (C), Specificity (S), and Verification (V) across all prompts.
\captionsetup[table]{justification=centering}

\begin{table}[htbp]
\vspace{-10pt}
\centering

\begin{minipage}{0.45\textwidth}
\centering
\small

\label{tab:prompt-structure}
\begin{tabular}{lrrr}
\toprule
 & C & S & V \\
\midrule
Mean & 1.48 & 1.08 & 0.58 \\
Std. Dev. & 0.62 & 0.58 & 0.57 \\
\bottomrule
\end{tabular}

\vspace{1mm}
{\footnotesize (a) Prompt structure (C, S, V)}
\end{minipage}
\hfill
\begin{minipage}{0.45\textwidth}
\centering
\small
\begin{tabular}{lccc}
\toprule
Class & Mean & Median & N \\
\midrule
PA & 3.96 & 4 & 89 \\
PN & 3.00 & 3 & 53 \\
NE & 2.11 & 2 & 80 \\
CL & 3.53 & 4 & 43 \\
\bottomrule
\end{tabular}

{\footnotesize (b) PQS by outcome class}
\end{minipage}

\vspace{-6pt}
\caption{Prompt structure dimensions and PQS by outcome class.}
\vspace{-15pt}
\end{table}

\begin{figure}[htbp]
\vspace{-10pt}
\centering
\includegraphics[width=.7\textwidth]{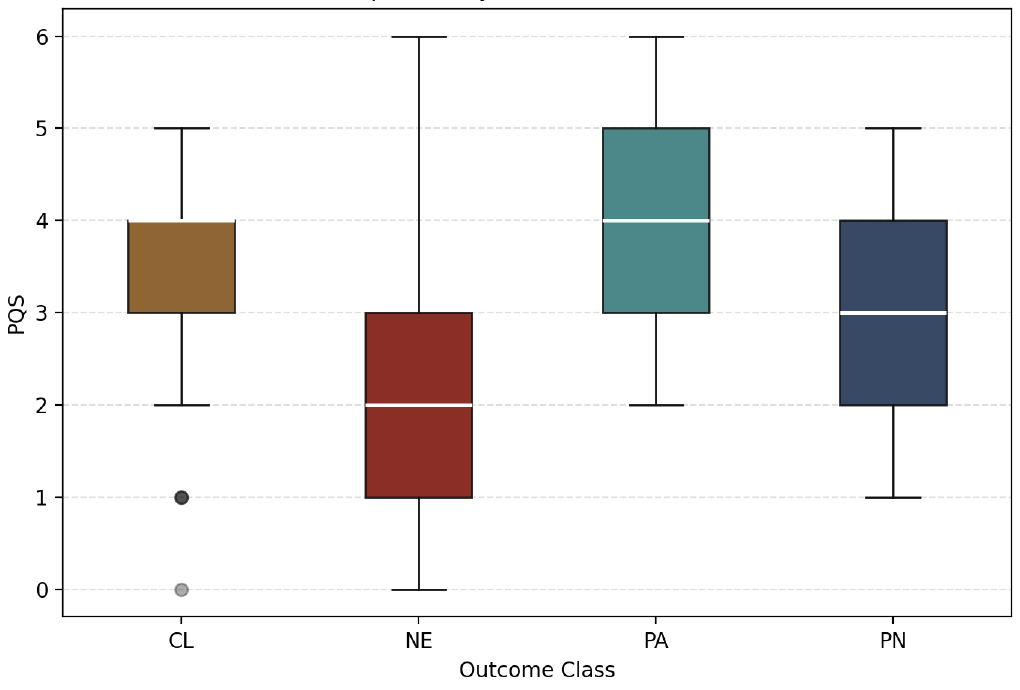}
\vspace{-16pt}
\caption{Distribution of Prompt Quality Score (PQS) across outcome classes.}
\label{fig:pqs_boxplot}
\vspace{-10pt}
\end{figure}

Prompts exhibit higher levels of Context (mean = 1.48) and
Specificity (mean = 1.08) than Verification (mean = 0.58),
indicating that prompts more frequently provide contextual grounding
and task definition than explicit correctness signals. Prompt quality
also varies substantially across outcome classes. PA cases exhibit
the highest mean PQS (3.96), whereas NE cases exhibit the lowest
(2.11), suggesting a strong association between prompt quality and
successful code generation. PN and CL occupy intermediate ranges with
substantial overlap, indicating that mid-range prompt quality alone
does not uniquely determine outcomes, as also reflected in the
overlapping distributions shown in Figure~\ref{fig:pqs_boxplot}.
Notably, CL cases exhibit relatively high PQS despite not being
merged, suggesting that prompt structure is more closely associated
with early-stage outcomes such as code generation and adoption than
with downstream pull-request lifecycle outcomes.

Pull request size and contributor experience exhibit highly
right-skewed distributions, with large discrepancies between mean and
median values and the presence of extreme outliers. These
characteristics motivate the use of log transformations and robust
modeling techniques in subsequent analyses. Together, these
descriptive patterns suggest that the influence of prompt structure
evolves across the pull request workflow, motivating the stage-based
modeling framework introduced next, which decomposes the workflow
into code generation, adoption, integration, and lifecycle phases.

\subsection{Stage-Based Modeling of Prompt Structure Effects}
\label{sec:gate_modeling}

To examine how prompt structure influences progression through the
pull request workflow, we model the process as a sequence of stages
corresponding to code generation, adoption, integration, and
lifecycle outcomes. This decomposition isolates how different
prompt dimensions influence each transition rather than treating
pull request outcomes as a single aggregated process.

We report results for each stage independently, beginning with
code generation (Gate~0), followed by adoption (Gate~1),
integration depth (Gate~2), and lifecycle outcomes. For each
stage, we estimate regression models that quantify the effects of
Context, Specificity, and Verification while controlling for
pull request characteristics.

\vspace{5pt}
\noindent \textbf{Qualitative interpretation and illustrative examples.}
To complement the quantitative modeling, we use the directed
qualitative analysis described in Section~\ref{sec:method-qual} to
identify recurring patterns in prompt formulation, implementation
framing, evaluability cues, and downstream integration behavior. For
each workflow stage, we summarize these patterns and present
purposefully sampled representative and contrasting examples spanning
different prompt-quality configurations, workflow outcomes, and
technical contexts. These examples are intended to illustrate
recurring patterns identified during the qualitative review rather
than serve as standalone evidence or estimates of prevalence.
Additional examples are included in the replication package, and each
case is labeled as \textit{PA-X}, \textit{PN-X}, or \textit{NE-X} to
support cross-referencing.

\subsubsection{Gate 0: Code Generation}
\label{sec:gate0_results}

\noindent Gate~0 examines whether prompt structure influences
whether ChatGPT produces actionable code. This stage distinguishes
cases where code was generated (PA and PN) from interactions that
produced only textual or conceptual guidance (NE). After excluding
closed pull requests (CL) and observations with missing values,
the effective sample size is $n=218$.
Table~\ref{tab:gate0_results} reports odds ratios (OR) and 95\%
confidence intervals from logistic regression models predicting
code generation. Model~1 includes the three prompt dimensions
(Context, Specificity, and Verification), while Model~2 adds
log-transformed pull request size as a control variable.

\begin{table}[h]
\centering
\vspace{-4pt}
\caption{Gate 0: Logistic Regression Results for Code Generation}
\vspace{-6pt}
\label{tab:gate0_results}

\begin{tabular}{lcc}
\toprule
 & \textbf{(1) Baseline} & \textbf{(2) + PR Size} \\
\midrule
Context (C)
& 2.16*  & 2.14*  \\
& [1.14, 4.10] & [1.13, 4.07] \\

Specificity (S) 
& 62.96***  & 65.84***  \\
&[8.52, 465.19] & [8.90, 487.14] \\

Verification (V)
& 0.90  & 0.90  \\
&[0.43, 1.88] & [0.43, 1.88] \\

Log(PR Size) 
& ---  & 1.12 \\
& ---  & [0.91, 1.36] \\

\midrule
Observations & 218 & 218 \\
\bottomrule
\multicolumn{3}{l}{\footnotesize Notes: Odds ratios reported with 95\% confidence intervals in brackets.} \\
\multicolumn{3}{l}{\footnotesize *** $p < 0.001$, ** $p < 0.01$, * $p < 0.05$.} \\
\end{tabular}

\vspace{-10pt}
\end{table}

\noindent Both \textit{Context} and \textit{Specificity} are strong predictors of actionable code generation. A one-unit increase in Context is associated with approximately a twofold increase in the odds of generating actionable code (OR $\approx 2.1$), and this effect remains statistically significant across model specifications. \textit{Specificity} exhibits the strongest association with code generation (OR $\approx 66$), indicating that prompts with clearer goals, constraints, and implementation requirements are substantially more likely to produce actionable code. Although the estimated effect size is large, the corresponding confidence interval is also wide, and its precise magnitude should therefore be interpreted cautiously.

In contrast, \textit{Verification} does not exhibit a statistically significant effect, suggesting that correctness-related cues do not determine whether code is generated. Adding pull request size does not materially change the estimates, and PR size itself is not statistically significant. Overall, these findings suggest that code generation depends primarily on prompt grounding and task specification, whereas evaluative and correctness-oriented cues appear more relevant to downstream workflow stages.

\vspace{5pt}
\noindent \textbf{Recurring structural patterns.}
Across the reviewed Gate~0 cases, recurring structural differences
emerged between prompts that produced actionable code and those that
did not. Prompts associated with actionable generation frequently
framed bounded implementation objectives, specified concrete
transformation or construction tasks, or constrained the expected
output format. In contrast, prompts associated with non-code outcomes
were often exploratory, conceptual, or discussion-oriented despite
being grounded in valid technical contexts. These patterns suggest that actionable code generation is most
commonly associated with prompts that frame bounded implementation
tasks rather than exploratory or discussion-oriented requests.

\vspace{5pt}
\noindent \textbf{Illustrative examples.}
To further interpret these effects, we examine representative PN and
NE cases reflecting the recurring structural patterns identified during
the qualitative review. 
The examples contrast implementation-oriented and conceptual prompt
formulations to illustrate how prompt structure influences code
generation outcomes.

\noindent \faCheckSquareO \textbf{Implementation-oriented prompts enabling code generation.}
Case~PN-19 (C=1, S=1, V=1) illustrates how implementation-oriented
prompts can generate actionable code even when the generated solution
is not ultimately adopted. The developer asks:
\textit{``Hello can you give me a regex to match ULID format?''}
and later evaluates a candidate expression:
\textit{``This one is correct? [0-7][0-9A-HJKMNP-TV-Z]\{25\}''}.
Although the prompt provides only moderate contextual grounding, it
defines a clear implementation target and bounded coding task,
enabling ChatGPT to generate a usable regex solution. This example
illustrates why Context and Specificity strongly influence code
generation at Gate~0.

\noindent \faCheckSquareO \textbf{Conceptual prompts limiting code generation.}
By contrast, Case~NE-3 (C=1, S=0, V=0) asks:
\textit{``how are express.js middleware functions commonly named? [...]
What might you call a middleware that adds a content security policy
[...]?''}. Although the prompt is grounded in a concrete technical
domain, it does not define a bounded implementation task or request
an executable artifact. As a result, ChatGPT produces descriptive
guidance rather than actionable code. This example illustrates that
contextual grounding alone is insufficient to support code generation
when prompts remain conceptual rather than implementation-oriented.

\begin{custombox}
{\faLightbulbO \hspace{0.01in} \textbf{Summary of Gate 0 Results:}}
Code generation is driven primarily by \textit{Specificity} and
\textit{Context}. Prompts that clearly define bounded implementation
objectives and provide relevant technical grounding are substantially
more likely to produce actionable code, whereas \textit{Verification}
does not significantly influence whether code is generated. The
qualitative analysis reinforces this pattern by showing that
implementation-oriented prompts tend to yield executable solutions,
while conceptual or exploratory prompts more often produce
explanatory guidance.
\end{custombox}

\subsubsection{Gate 1: Code Adoption}
\label{sec:gate1_results}

\noindent Gate~1 examines whether prompt structure influences
whether generated code is adopted within the pull request. This
analysis is restricted to cases where code was generated (PA and PN
outcomes). After removing observations with missing values, the
effective sample size is $n=141$.

Table~\ref{tab:gate1_results} reports odds ratios (OR) and 95\%
confidence intervals from logistic regression models predicting code
adoption. Model~1 includes Context, Specificity, and Verification,
while Model~2 adds log-transformed pull request size as a control
variable.

\begin{table}[h]
\centering
\vspace{-4pt}
\caption{Gate 1: Logistic Regression Results for Code Adoption}
\vspace{-6pt}
\label{tab:gate1_results}

\begin{tabular}{lcc}
\toprule
 & \textbf{(1) Baseline} & \textbf{(2) + PR Size} \\
\midrule
Context (C) 
& 2.28*& 2.22 \\
& [1.02, 5.10] & [0.97, 5.09] \\

Specificity (S) 
&  0.61  & 0.79 \\
& [0.23, 1.58] & [0.29, 2.15] \\

Verification (V) 
& 7.78*** & 8.45***  \\
& [3.40, 17.82] & [3.50, 20.43] \\

Log(PR Size) 
& --- &  1.36**  \\
& --- & [1.08, 1.71] \\

\midrule
Observations & 141 & 141 \\
\bottomrule
\multicolumn{3}{l}{\footnotesize Notes: Odds ratios reported with 95\% confidence intervals in brackets.} \\
\multicolumn{3}{l}{\footnotesize *** $p < 0.001$, ** $p < 0.01$, * $p < 0.05$.} \\
\end{tabular}

\vspace{-10pt}
\end{table}

\noindent \textit{Verification} is the strongest and most consistent predictor of code adoption. A one-unit increase in Verification is associated with approximately eightfold higher odds of patch adoption (OR $\approx 7.8$--$8.5$), and this effect remains highly significant across model specifications. In contrast, \textit{Specificity} does not exhibit a statistically significant association with adoption, indicating that clearly specifying task requirements alone is insufficient to distinguish adopted from non-adopted patches once actionable code has been generated. \textit{Context} shows a positive association with adoption in the baseline model, but the effect becomes marginal after controlling for pull request size, suggesting a more limited role at this stage of the workflow.

Including pull request size reveals a significant positive effect, with larger pull requests more likely to incorporate LLM-generated code. Importantly, adding this control does not materially change the estimated effect of Verification, indicating that the observed relationship is robust to differences in pull request complexity.

Overall, these results suggest that adoption depends primarily on
whether generated outputs can be evaluated against explicit
correctness expectations. Whereas Context and Specificity help
developers communicate and constrain implementation tasks during
code generation, Verification provides validation criteria and
evaluative signals that appear to increase confidence in generated
solutions and their suitability for integration. This contrast
highlights a transition from task formulation during generation to
correctness assessment during adoption.

\vspace{5pt}
\noindent \textbf{Recurring structural patterns.}
The qualitative analysis revealed recurring prompt patterns
associated with both successful and unsuccessful adoption outcomes.
Prompts linked to adopted code frequently included explicit
correctness criteria, observable output expectations, or other
evaluative signals that enabled developers to assess generated
solutions with confidence. In contrast, prompts associated with
non-adopted code often produced technically plausible outputs and
sometimes provided substantial contextual grounding and task
specification, yet lacked sufficiently explicit validation criteria
or acceptance conditions to support confident evaluation within the
target implementation context. These recurring patterns
qualitatively reinforce the quantitative finding that Verification
plays a central role during adoption, whereas Context and
Specificity alone are often insufficient to explain adoption
decisions.

\vspace{5pt}
\noindent \textbf{Illustrative examples.}
To further interpret these effects, we examine representative PA and
PN cases reflecting the recurring structural patterns identified during
the qualitative review. The examples illustrate how adoption depends
less on contextual grounding alone and more on whether generated
outputs are sufficiently constrained and evaluable for developers to
assess correctness with confidence. Some examples intentionally
reappear across gates to illustrate how the same prompt may exhibit
different strengths and limitations across successive workflow stages.

\noindent \faCheckSquareO \textbf{Explicit constraints and observable correctness enabling adoption.}
Case~PA-22 (C=1, S=2, V=2) illustrates how high Specificity and
Verification support direct adoption of generated code. The
developer asks:
\textit{``Please generate code using the Kotlin \texttt{sksamuel/scrimage}
library to draw text on an image and add an outline to the text. The
text should be white and the outline black.''}. Although the prompt
provides only moderate contextual grounding, it defines a narrowly
scoped implementation objective together with explicit output
constraints. The prompt also includes observable correctness
conditions, namely that the generated image must contain white text
with a black outline, making the generated solution directly
evaluable and reducing ambiguity regarding expected behavior. This
example illustrates why Specificity and Verification emerge as
significant predictors of adoption at Gate~1.

\noindent \faCheckSquareO \textbf{Generated but weakly evaluable code limiting adoption.}
By contrast, Case~PN-3 (C=2, S=2, V=0) illustrates how code may fail
to be adopted despite strong contextual grounding and task
specification. The developer asks whether contextual information
normally managed through SLF4J's MDC can instead be attached directly
through a fluent \texttt{log.atLevel()} logging approach in a highly
concurrent environment and requests an example Log4j2 configuration.
Although the prompt provides substantial technical context and a
clearly defined objective, it does not specify expected behavior,
acceptance criteria, or other correctness conditions that would allow
the generated solution to be evaluated within the target system. This
example illustrates that adoption depends not only on obtaining
plausible code or configuration guidance, but also on whether prompts
provide explicit evaluative signals that support trustworthy
assessment.

\begin{custombox}
{\faLightbulbO \hspace{0.01in} \textbf{Summary of Gate 1 Results:}}
Gate~1 results show that code adoption is driven primarily by
\textit{Verification}. Prompts containing explicit correctness
expectations or validation criteria are substantially more likely to
produce code that developers adopt, whereas \textit{Specificity} does
not distinguish adopted from non-adopted code once generation has
occurred. \textit{Context} plays a weaker role and becomes marginal
after accounting for pull request size. The qualitative analysis
supports this finding by showing that adoption depends less on task
formulation and more on whether generated outputs can be evaluated
against clear correctness expectations. Together, these results
suggest that Gate~1 represents a transition from code generation to
solution evaluation.
\end{custombox}

\subsubsection{Gate 2: Integration Depth}
\label{sec:gate2_results}

\noindent Gate~2 examines whether prompt structure influences the
extent to which LLM-generated code is integrated into the final
implementation. This analysis is restricted to adopted cases (PA
outcomes). After removing observations with missing values, the
effective sample size is $n=89$, spanning 71 repositories. The
dependent variable, $\textit{fraction\_adopted} \in (0,1]$,
represents the proportion of generated code incorporated into the
final implementation.

\begin{table}[h]
\centering
\vspace{-4pt}
\caption{Gate 2: Fractional Logit Results for Fraction of Code Adopted}
\vspace{-6pt}
\label{tab:gate2_results}

\begin{tabular}{lc}
\toprule
Variable & AME \\
\midrule
Context (C) 
& 0.125* \\
& [0.016, 0.233] \\

Specificity (S) 
& -0.013 \\
& [-0.131, 0.105] \\

Verification (V) 
& -0.008  \\
& [-0.106, 0.089] \\

Log(PR Size) 
&  -0.036  \\
& [-0.074, 0.002] \\

\midrule
Observations & 89 \\
\bottomrule
\multicolumn{2}{l}{\footnotesize Notes: Average marginal effects reported with 95\% confidence intervals in brackets.} \\
\multicolumn{2}{l}{\footnotesize Standard errors are clustered at the repository level.} \\
\multicolumn{2}{l}{\footnotesize *** $p < 0.001$, ** $p < 0.01$, * $p < 0.05$.} \\
\end{tabular}

\vspace{-10pt}
\end{table}

Table~\ref{tab:gate2_results} reports average marginal effects (AME)
from the fractional-logit model with repository-clustered robust
standard errors.
\textit{Context} is the only statistically significant predictor of
integration depth. A one-unit increase in Context is associated with
an increase of approximately 12.5 percentage points in the expected
fraction of adopted code. In contrast, \textit{Specificity} and
\textit{Verification} do not exhibit statistically significant
effects, indicating that while these dimensions influence earlier
workflow stages, they do not determine how extensively code is reused
once accepted. Log(PR Size) shows a negative but statistically
insignificant association, and including this control does not
materially affect the estimated impact of Context.

Overall, these results suggest that deeper integration depends
primarily on contextual fit. Once generated code has been accepted,
the extent of reuse appears to depend less on task specification or
correctness cues and more on how well the generated solution aligns
with the surrounding implementation environment.

\vspace{5pt}
\noindent \textbf{Recurring structural patterns.}
The qualitative analysis revealed recurring prompt patterns associated
with different levels of downstream reuse after adoption. Prompts
linked to deeper integration frequently embedded generated tasks within
rich implementation context, including surrounding code structures,
existing APIs, configuration details, or repository-specific logic.
This contextual grounding enabled generated outputs to align more
closely with the surrounding implementation environment and therefore
required less downstream adaptation before reuse. In contrast, prompts
with high Specificity but limited contextual grounding often produced
technically correct yet relatively generic outputs that required
substantial refinement or restructuring prior to integration. 
These recurring patterns qualitatively reinforce the finding that
deeper reuse depends primarily on contextual fit rather than task
specification or correctness cues alone.

\vspace{5pt}
\noindent \textbf{Illustrative examples.}
To further interpret these effects, we examine representative PA
cases reflecting recurring structural patterns observed during the
qualitative review. The examples illustrate how deeper integration
depends primarily on contextual grounding and the extent to which
generated solutions align with existing implementation structures
and surrounding code artifacts. Some examples intentionally reappear
across gates to illustrate how the same prompt may exhibit different
strengths and limitations across successive workflow stages.

\noindent \faCheckSquareO \textbf{Strong contextual grounding enabling deep integration.}
Case~PA-78 (C=2, S=2, V=1; Fraction=84.84\%) illustrates how strong
contextual grounding enables extensive reuse after adoption. The
developer provides a complete Python program and asks:
\textit{``Refactor the following code to improve readability and structure:
[...]''}. The prompt includes multiple function definitions,
control-flow structures, and surrounding implementation details,
clearly situating the task within an existing codebase. This rich
Context enables the generated solution to align closely with the
surrounding implementation, allowing most of the output to be
incorporated with minimal modification. This example illustrates
why Context emerges as the dominant predictor of integration depth
at Gate~2.

\noindent \faCheckSquareO \textbf{High specificity with limited contextual grounding.}
By contrast, Case~PA-24 (C=1, S=2, V=1; Fraction=7.69\%)
illustrates how highly specific implementation requests may still
result in limited reuse when contextual grounding is weak. The
developer asks:
\textit{``Create a TypeScript enum for the following values and initialize
the corresponding variables [...]''}. Although the prompt clearly
defines the intended structure and functionality, it provides
little information about the surrounding implementation
environment or existing code structures. As a result, the generated
output remains relatively generic and requires substantial
downstream adaptation before incorporation into the pull request.
This example illustrates why contextual alignment, rather than
Specificity alone, drives deeper reuse at Gate~2.

\begin{custombox}
{\faLightbulbO \hspace{0.01in} \textbf{Summary of Gate 2 Results:}}
Gate~2 results show that integration depth is driven primarily by
\textit{Context}. Prompts containing rich implementation context are
associated with deeper reuse of generated code, whereas
\textit{Specificity} and \textit{Verification} do not significantly
influence how extensively accepted code is incorporated into the final
implementation. The qualitative analysis reinforces this finding by
showing that generated solutions are reused more extensively when they
align closely with existing code structures, APIs, and surrounding
implementation details. Together, these results suggest that Gate~2 is
primarily a problem of implementation alignment rather than task
specification or correctness evaluation.
\end{custombox}

\subsection{Axis B: PR Lifecycle Results}
\label{sec:axisb_results}

\noindent Axis~B models pull request lifecycle as a time-to-event
process, capturing the rate at which pull requests are merged or
closed without merge. After removing observations with missing
values, the effective sample size is $n=261$. Pull requests resolve
via either merge or closure, with unresolved cases treated as
right-censored observations.

Table~\ref{tab:axisb_results} reports hazard ratios (HR) and 95\%
confidence intervals from cause-specific Cox proportional hazards
models for merge and close events. None of the prompt dimensions
(Context, Specificity, or Verification) exhibit statistically
significant effects in the merge model, indicating that prompt
structure does not substantially influence the rate at which pull
requests are merged. In contrast, \textit{Verification} is a
significant predictor in the close-hazard model
(HR $\approx 1.77$), indicating that higher levels of verification
are associated with faster closure without merge. Context and
Specificity do not exhibit significant effects in this model.

\begin{table}[h]
\centering
\vspace{-4pt}
\caption{Axis B: Cause-Specific Cox Results for PR Lifecycle}
\vspace{-6pt}
\label{tab:axisb_results}

\begin{tabular}{lcc}
\toprule
 & \textbf{Close Hazard} & \textbf{Merge Hazard} \\
\midrule
Context (C) 
& 1.72  &  1.02 \\
& [0.95, 3.13] &  [0.80, 1.30] \\

Specificity (S) 
&  0.63 &  1.07  \\
&  [0.34, 1.16] & [0.81, 1.41] \\

Verification (V) 
& 1.77*  & 0.84  \\
&[1.03, 3.05] & [0.65, 1.09] \\

Log(PR Size) 
& 0.83**  & 0.93*  \\
& [0.72, 0.95] & [0.87, 1.00] \\

\midrule
Observations & 261 & 261 \\
\bottomrule
\multicolumn{3}{l}{\footnotesize Notes: Hazard ratios reported with 95\% confidence intervals in brackets.} \\
\multicolumn{3}{l}{\footnotesize *** $p < 0.001$, ** $p < 0.01$, * $p < 0.05$.}\\ 
\end{tabular}
\vspace{-10pt}
\end{table}

\noindent Log(PR size) is significant in both models, with hazard
ratios below one for merge (HR $\approx 0.93$) and closure
(HR $\approx 0.83$), indicating that larger pull requests resolve
more slowly overall. The stronger effect for closure suggests that
larger pull requests are less likely to be quickly rejected or
abandoned, but also require more time to reach resolution.

Overall, these results suggest that lifecycle dynamics are driven
by factors distinct from those influencing earlier workflow stages.
While prompt structure strongly affects code generation, adoption,
and integration depth (Gates~0--2), it does not appear to
substantially influence merge decisions once code has entered the
pull request workflow. Instead, \textit{Verification} is associated
primarily with faster negative resolution outcomes. One possible
explanation is that prompts emphasizing correctness constraints make
implementation mismatches or unmet requirements easier to identify
during review, allowing maintainers to reach closure decisions more
quickly when generated solutions do not align with project
expectations. More broadly, lifecycle outcomes appear to reflect
repository-level coordination and review processes that extend
beyond prompt design.

\begin{custombox}
{\faLightbulbO \hspace{0.01in} \textbf{Summary of Axis B Results:}}
Axis~B results show that prompt structure does not significantly
influence the rate at which pull requests are merged, although
verification signals are associated with faster closure without
merge. In contrast to earlier workflow stages, life cycle timing is
driven primarily by pull request size and repository-level review
processes. These findings suggest that prompt design has limited
influence on pull request life cycle dynamics, reflecting a transition
from LLM-driven effects to broader coordination and review factors.
\end{custombox}

\subsection{Robustness and Sensitivity Analysis}
\label{sec:robustness-results}

We conducted a series of robustness and sensitivity analyses to assess
whether the stage-dependent findings were sensitive to repository
heterogeneity, dominant repositories or programming languages,
skewed pull request distributions, alternative Gate~2
operationalizations, prompt-length effects, and model-specification
choices. Repository-level robustness checks using
repository-clustered robust standard errors and repository
random-intercept sensitivity specifications preserved the primary
findings across all gates. Gate~0 continued to show positive effects
for \textit{Context} and \textit{Specificity},
Gate~1 consistently
identified \textit{Verification} as the primary predictor of code
adoption, and Gate~2 continued to identify \textit{Context} as the
primary predictor of integration depth.

Additional sensitivity analyses excluding dominant repositories,
frequent programming languages, and extreme pull request outliers
did not materially alter the qualitative interpretation of the
Gate~0 and Gate~1 models. Gate~2 \textit{Context} effects remained
directionally positive across restricted samples, although they
became less stable under smaller PA-only subsets, indicating
greater sensitivity at the integration-depth stage. Alternative
Gate~2 operationalizations based on ordinal reuse levels similarly
identified \textit{Context} as the strongest positive predictor,
while \textit{Specificity} and \textit{Verification} remained
comparatively weak. Models using the aggregate Prompt Quality Score
(PQS) positively predicted Gate~0 and Gate~1 outcomes, but did not
consistently predict Gate~2 integration depth, reinforcing the
importance of analyzing prompt dimensions separately for deeper
reuse behaviors.

Finally, introducing prompt-length controls did not materially
alter the primary interpretations, suggesting that the observed
\textit{Context} effects were not simply artifacts of prompt
verbosity. Stratified 80/20 hold-out stability checks also
preserved the principal coefficient directions, although several
effects weakened under smaller hold-out subsets, particularly for
Gate~2 and other restricted PA-only analyses. Overall, the
robustness analyses support the interpretation that the observed
stage-dependent relationships reflect stable characteristics of
developer--LLM interaction behavior rather than artifacts of a
particular repository subset, language distribution, or modeling
specification.

\subsection{Model Diagnostics}
\label{sec:model-diagnostics}

To evaluate the validity of the statistical models, we conducted a
series of diagnostic checks across all specifications. Variance
inflation factors (VIF) remained below 2 for all predictors,
indicating no evidence of problematic multicollinearity and
suggesting that \textit{Context}, \textit{Specificity},
\textit{Verification}, and PR size capture distinct aspects of
prompt structure and workflow behavior.

All logistic regression models associated with Gates~0 and~1
converged successfully within a small number of iterations.
Similarly, the fractional
logit models used for Gate~2 converged without numerical
instability, and fitted values aligned closely with the observed
integration-depth distributions, supporting the suitability of the
modeling framework for bounded reuse outcomes.

For Axis~B, proportional hazards diagnostics based on Schoenfeld
residuals indicated mild deviations from the proportional hazards
assumption for certain predictors, particularly PR size.
Accordingly, the reported hazard ratios should be interpreted as
approximate relative effects rather than strictly time-invariant
estimates. Overall, the diagnostic analyses did not identify severe
violations of model assumptions, supporting the validity and
stability of the reported results.

\subsection{Overall Findings and Discussion (RQ1--RQ2)}
\label{sec:overall-discussion}

Our study examined how prompt structure relates to different stages of
LLM-assisted pull request workflows, ranging from annotation
reliability and code generation to adoption and integration depth.
Across the analyzed pull requests, prompts were associated with a wide
range of development activities, including implementation requests,
debugging, refactoring, architectural reasoning, configuration
changes, and conceptual discussion. Rather than treating LLM usage as
a binary notion of success or failure, our analysis models
AI-assisted development as a staged collaborative workflow in which
different prompt characteristics become important at different stages
of developer interaction, evaluation, and integration.

These findings complement prior studies of self-admitted
generative-AI usage in open-source projects, which show that
developers increasingly share ChatGPT interactions in issues and pull
requests for tasks such as code generation, debugging, explanation,
and design discussion~\cite{tufano2026developers}. Whereas such work
primarily characterizes the breadth of developer GenAI usage, our
study examines what happens after these interactions enter
collaborative pull request workflows: whether prompts generate
actionable code, whether developers adopt the generated code, and how
deeply the generated code is integrated into the surrounding
implementation context.

Synthesizing findings across RQ1 and RQ2, we identify several
cross-cutting patterns that clarify how developers structure prompts
and evaluate LLM-generated outputs during collaborative software
development.

\vspace{5pt}
\noindent \textbf{1. Reliable prompt annotation requires selective human oversight.}
RQ1 shows that annotation agreement and automation suitability are not
identical. Although \textit{Context} achieves the strongest overall
human--LLM agreement, it also exhibits systematic under-scoring bias,
making fully automated scoring unreliable under our evaluation
criteria. In contrast, \textit{Specificity} exhibits lower aggregate
agreement but more stable scoring behavior across several workflow
contexts, allowing selective automation in structurally explicit
contexts. \textit{Verification} remains unstable when correctness
signals are implicit or distributed across interactions.

Agreement further varies across workflow outcomes. Conceptual and descriptive interactions often contain more localized
structural signals, whereas iterative code-centric exchanges may
involve more distributed context and implicit evaluative reasoning,
reducing annotation stability. These findings suggest that reliable operationalization of prompt
structure requires class-aware hybrid human--LLM workflows rather
than uniform automation. More broadly, they demonstrate that the
reliability of LLM-assisted annotation depends both on the construct
being measured and the workflow context in which it is observed.

These findings align with a growing body of research showing that
LLM-assisted annotation workflows require selective human oversight
rather than fully automated labeling
pipelines~\cite{wang:CHI:2024,
Takehi:SIGIR:2025,
Xu:LAK26:2026}. Prior work highlights how annotation reliability may
vary depending on task complexity, verification difficulty,
contextual ambiguity, and interaction design. 
Our results extend these observations to developer--LLM interactions
by showing that annotation reliability depends not only on the target
construct, but also on conversational context and the distribution of
evaluative reasoning across interactions. Consequently, for empirical
software-engineering studies that rely on LLM-based annotation,
validation should be performed at the level of individual constructs
and workflow contexts rather than assuming uniform annotation
reliability across a dataset.

\noindent \textbf{2. Prompt effectiveness is workflow-stage-dependent.}
The results reveal a clear stage-dependent relationship between prompt
structure and workflow outcomes. Prompts that provide technical
grounding and bounded implementation objectives are more likely to
generate actionable code, whereas downstream adoption depends primarily on verification-oriented cues,
including correctness expectations, evaluability constraints, and
observable output criteria that support developer assessment and trust.
Deeper
integration, in turn, depends primarily on contextual alignment with
the surrounding implementation environment rather than task definition
alone.

These findings extend recent work on ChatGPT-assisted issue
resolution, which shows that conversational helpfulness depends on
task, project, and interaction
characteristics~\cite{Ehsani:EMSE:2026}. They also complement prior
work on AI-assisted pull request integration. \texttt{PatchTrack} showed that
ChatGPT-generated patches frequently undergo adaptation, selective
reuse, or complete rejection during collaborative review, indicating
that successful AI assistance cannot be understood solely through
generated-code quality or merge outcomes~\cite{ogenrwot2026patchtrack}.
Whereas \texttt{PatchTrack} focused on the downstream fate of generated
patches, our analysis investigates an earlier stage of the workflow by
examining how prompt structure influences actionable code generation,
code adoption, and integration depth. Together, the two studies
suggest that variation in downstream integration outcomes may partly
originate from upstream differences in prompt construction and
contextual grounding.

The findings further refine recent prompt-engineering results showing
that prompt effectiveness remains task-dependent even for advanced
LLMs~\cite{Wang:TOSEM:2025}. Our results suggest that prompt
effectiveness is also workflow-stage-dependent: the prompt
characteristics supporting code generation are not identical to those associated with downstream adoption and integration.
Specifically,
\textit{Specificity} and \textit{Context} are most strongly associated
with actionable code generation, \textit{Verification} emerges as the
primary predictor of code adoption, and \textit{Context} again becomes
the dominant factor for deeper integration. These findings suggest
that different dimensions of prompt quality become important at
different stages of AI-assisted software development.

\noindent \textbf{3. AI-assisted development reflects iterative evaluation, adaptation, and alignment.}
Across workflow stages, developers evaluate generated outputs using
traditional software engineering criteria such as implementation fit,
maintainability, architectural alignment, and project conventions.
LLM-generated suggestions are therefore incorporated into existing
review and collaboration processes rather than bypassing them. Prior
work further shows that repeated failures, poor contextual fit, and
unmet developer expectations can erode trust in LLM-generated
suggestions and ultimately lead developers to abandon AI assistance,
highlighting the importance of evaluation and alignment within
AI-assisted development workflows~\cite{Tie:TOSEM:2026}.

The findings further suggest that verification-oriented reasoning
becomes increasingly important during downstream evaluation and
adoption decisions. Prompts containing explicit correctness
expectations, evaluability constraints, or observable output criteria
are more likely to support developer trust and confident integration.
At the same time, qualitative analyses show that such evaluative
signals are often implicit or distributed across interactions,
helping explain why verification-oriented constructs remain more
difficult to operationalize reliably through automated annotation.

More broadly, these findings support recent arguments that prompts
should be treated as first-class software artifacts within
prompt-enabled systems~\cite{Chen:TOSEM:2026,
Villamizar:PROFES:2025} and align with qualitative evidence that
developers engage in iterative ``prompt programming''
practices~\cite{Liang:FSE:2025}. Our results extend this emerging
literature by showing how prompt structure relates not only to prompt
construction or conversational helpfulness, but also to downstream
collaborative software-engineering outcomes such as code adoption and
integration depth.

Viewed collectively, the findings suggest that AI-assisted software
development is best understood as a staged collaborative process
rather than a single generation event. In this view, prompt
construction, code generation, adoption, and integration form a
connected workflow in which upstream prompting decisions influence
downstream collaborative outcomes. This perspective complements prior
work on patch integration~\cite{ogenrwot2026patchtrack} by extending
the analysis further upstream to the prompts that initiate
AI-assisted development activities.
The results further demonstrate that studying these workflows requires
careful operationalization of prompt-quality constructs and selective
use of LLM-assisted annotation, linking methodological choices about
measurement directly to substantive conclusions about AI-assisted
software-development practices.

\vspace{5pt}
\noindent \textbf{Temporal scope and model evolution.}
The dataset spans a period during which ChatGPT and related LLMs underwent substantial capability improvements. Because historical conversations rarely disclose the exact model version used, we cannot directly assess the impact of model evolution on individual outcomes. Consequently, our findings should not be interpreted as measuring the performance of a specific ChatGPT release. Prior work suggests that although advances in LLM capabilities may reduce the effectiveness of some prompting strategies, prompt structure and task-specific guidance remain important for software-engineering tasks~\cite{Wang:TOSEM:2025}. Accordingly, our results should be interpreted as characterizing AI-assisted development workflows during the period represented by the dataset rather than the behavior of a particular model version.

\subsection{Implications}
\label{sec:implications}

\paragraph{Implications for Practitioners}

The findings suggest that effective AI-assisted software development
depends on structuring prompts differently across workflow stages.
During early-stage code generation, prompts benefit from clear
technical grounding and bounded implementation objectives, including
relevant frameworks, APIs, code snippets, runtime behavior, and
expected functionality. Once code has been generated, successful adoption depends primarily on
verification-oriented cues such as expected behavior, validation
examples, output constraints, acceptance criteria, and other signals
that help developers evaluate correctness and trustworthiness.

The results further show that deep integration of LLM-generated code
depends heavily on contextual alignment with the surrounding
implementation environment. Even highly specific prompts may produce
correct and adoptable code that nevertheless requires substantial
downstream adaptation if insufficient implementation context is
provided. Developers seeking reusable LLM-generated solutions should
therefore prompt from within the target architectural and repository
context rather than treating code generation as an isolated task.

These findings further suggest that many downstream integration
challenges may originate upstream during prompt construction.
Developers often focus on improving generated code after it is
produced, yet our results indicate that contextual grounding and task
framing influence whether generated solutions are actionable,
adoptable, and ultimately reusable. Organizations seeking to improve
the effectiveness of AI-assisted development should therefore treat
prompt design as part of the software-development process rather than
as a separate interaction with an AI tool.

Beyond workflow outcomes, the findings demonstrate that organizations
adopting LLM-assisted annotation or evaluation pipelines should employ
selective human verification rather than uniform automation.
Annotation reliability varies substantially across prompt dimensions
and interaction contexts, indicating that human effort should be
concentrated where construct ambiguity, contextual reasoning, or
evaluative judgment remain difficult for LLMs to reproduce reliably.
Although \textit{Specificity} can be selectively automated in several
workflow contexts, \textit{Context} requires continued human oversight
due to systematic under-scoring bias, and \textit{Verification}
remains less stable when evaluative reasoning is implicit or
distributed across interactions. Collectively, these findings suggest that different dimensions of
prompt quality become important at different stages of AI-assisted
development: task definition and contextual grounding support code
generation, verification-oriented cues support adoption, and
contextual alignment becomes increasingly important for deeper
integration.
More broadly, the findings reinforce
that LLM-generated outputs should be treated as provisional artifacts
requiring critical evaluation, refinement, and adaptation within
existing software engineering workflows rather than as authoritative
solutions. This interpretation is consistent with recent evidence that
developers frequently reassess, revise, or abandon LLM assistance when
generated outputs fail to satisfy contextual requirements or maintain
developer trust~\cite{Tie:TOSEM:2026}.

\paragraph{Implications for Researchers}

This study highlights important methodological considerations for
future research on AI-assisted software engineering. First, prompt-quality
constructs such as Context, Specificity, and Verification cannot be
assumed to be uniformly observable or reliably annotatable across all
workflow settings. Second, LLM-assisted annotation reliability depends
both on the construct being measured and the workflow context in which
it is applied, motivating construct-aware and context-aware validation
strategies.

More broadly, the findings suggest that understanding AI-assisted
software development requires linking upstream prompting behavior to
downstream integration outcomes. Prior studies have examined either
developer--LLM interactions or the fate of generated artifacts in
collaborative workflows. Our results indicate that these perspectives
should be studied jointly, as prompt construction may represent an
important source of variation in subsequent adoption, adaptation, and
integration behavior.

The findings further demonstrate that scalable prompt-structure
operationalization requires adaptive hybrid annotation strategies
rather than uniform LLM automation. Future research on LLM-assisted
empirical software engineering should therefore move beyond aggregate
agreement metrics and develop evaluation protocols that explicitly
account for construct-specific variation, contextual ambiguity,
systematic annotation bias, and the differing reliability of prompt
dimensions across workflow contexts.

From a methodological perspective, the results suggest that future empirical studies using
LLMs as annotators should report construct-level reliability,
systematic bias analyses, and workflow-specific validation procedures
rather than relying solely on aggregate agreement measures. Such
reporting would improve transparency, facilitate comparison across
studies, and help researchers identify contexts in which automated
annotation remains appropriate or requires additional human oversight.

Finally, the results suggest that future benchmark design,
prompt-engineering research, and evaluations of LLM effectiveness
should incorporate richer implementation context and repository-level
information while recognizing that different prompt dimensions become
important at different stages of the development workflow. Evaluations
based solely on code generation or correctness may overlook factors
that influence downstream adoption and integration.

\paragraph{Implications for Educators}

The findings suggest that AI-assisted software development should be
taught as a staged collaborative workflow rather than as a simple code
generation activity. Educational use of tools such as ChatGPT should
therefore emphasize not only prompt construction, but also critical
evaluation, adaptation, and integration of AI-generated outputs within
real development settings. Students should also learn that prompt construction influences not
only the quality of generated outputs, but also the effort required
for downstream evaluation, adaptation, and integration. Prompting
should therefore be taught as a software-engineering activity that
affects the entire lifecycle of AI-assisted development rather than
only the initial generation step.

The results further highlight the importance of incorporating human
oversight and evaluative reasoning into AI-assisted software
engineering education. Cases involving plausible but weakly evaluable
or poorly integrated solutions provide opportunities to teach
reviewer trust, implementation fit, maintainability, and collaborative
decision-making. More broadly, the findings support instructional
approaches that position generative AI as a collaborative development
aid requiring structured human judgment rather than as an autonomous
code-generation system.

These insights are directly applicable to project-based software
engineering courses where students increasingly incorporate
generative AI into collaborative development activities. For example,
they are currently being integrated into senior design instruction at
UNLV through industry-based case studies illustrating how developers
evaluate, adapt, and integrate AI-generated solutions within
collaborative pull request workflows.
\section{Related Work}

\subsection{LLMs in Software Engineering and Developer--LLM Interactions}

Large language model (LLM)-powered tools such as ChatGPT and GitHub
Copilot are increasingly integrated into software engineering
workflows, supporting tasks including code generation, debugging,
testing, documentation, and issue
resolution~\cite{fan2023llm4se,hou2024large}. Recent studies show
that developers frequently use conversational LLMs within
collaborative development environments and openly share these
interactions in GitHub issues and pull
requests~\cite{hao2024githubconversations,xiao2024generative}.
Ehsani et al.~\cite{Ehsani:EMSE:2026} analyzed developer--ChatGPT
conversations in GitHub issue threads and found that successful
issue-resolution interactions depend on conversational, project, and
issue-level characteristics. Complementary work identified
prompt-level knowledge gaps, including missing context, missing
specifications, and unclear instructions, as recurring factors
associated with ineffective LLM-guided issue
resolution~\cite{ehsani2025promptgaps}.

Recent work has also begun studying self-admitted GenAI usage (SAGU)
as a lens for understanding AI adoption in open-source software
development. Xiao et al.~\cite{Xiao:TSE:2026} analyzed more than
1,200 public disclosures of GenAI usage across GitHub repositories
and developed taxonomies of developer tasks, generated content, and
usage purposes. Their findings establish self-admitted GenAI usage as
a valuable empirical artifact for studying developer--LLM
interactions, but focus primarily on adoption and disclosure behavior
rather than downstream integration outcomes.

A complementary line of research has examined failure modes in
developer--LLM interactions. Tie et al.~\cite{Tie:TOSEM:2026}
identified multiple categories of LLM failures, including incorrect
responses, context loss, and cognitive overload, and found that
persistent failures can erode developer trust and ultimately lead to
abandonment of AI assistance. Whereas their work focuses on why
developers disengage from LLM assistance, our study investigates how
prompt structure influences successful progression through subsequent
workflow stages, including code generation, adoption, and integration.

Beyond individual developer--LLM interactions, recent research has
examined how generative AI influences collaborative software
engineering workflows. Tufano et al.~\cite{tufano2026developers}
showed that developers use ChatGPT and GitHub Copilot for activities
extending beyond code generation, including debugging, testing,
documentation, refactoring, and code review. Similarly,
\texttt{PatchTrack} showed that ChatGPT-generated patches frequently
undergo adaptation, selective reuse, or complete rejection during
pull request integration, highlighting the socio-technical factors
that shape AI-assisted development
outcomes~\cite{ogenrwot2026patchtrack}. Together, these studies
highlight the importance of understanding how developer--LLM
interactions translate into downstream collaborative outcomes.

Whereas prior work primarily focuses on conversational helpfulness,
adoption behavior, abandonment, or the downstream fate of generated
artifacts, the present study investigates an earlier stage of the
workflow by examining how prompt structure influences actionable code
generation, code adoption, and integration depth across AI-assisted
pull request workflows.

\subsection{Prompting Practices and Prompt Quality}

Recent research increasingly highlights the importance of prompt
construction in AI-assisted software engineering. Ehsani et
al.~\cite{ehsani2025promptgaps} identified recurring deficiencies in
developer--ChatGPT interactions, including missing context,
underspecified requirements, competing objectives, and unclear
instructions, and proposed prompt-improvement heuristics emphasizing
specificity, contextual richness, and clarity. Similarly, Otten et
al.~\cite{otten2025promptinginpractice} found that developers
frequently rely on iterative multi-turn prompting and adapt their
prompts according to the software engineering task being performed.
Midolo et al.~\cite{midolo2026promptguidelines} further derived
practical prompting guidelines for code generation, emphasizing
explicit input/output specifications, examples, preconditions,
postconditions, and ambiguity clarification.

Recent studies have also examined how prompt strategies influence
software-engineering outcomes. Wang et
al.~\cite{Wang:TOSEM:2025} found that several prompt-engineering
techniques developed for earlier LLMs provide diminished benefits for
advanced models, whereas execution feedback and task-specific guidance
remain important for complex software-engineering tasks. Porta et
al.~\cite{Porta:EASE:2025} similarly reported that common prompt
patterns do not uniformly improve maintainability, security, or
reliability outcomes, suggesting that prompt effectiveness remains
strongly task- and context-dependent.

A complementary line of work argues that prompts should be treated as
software-engineering artifacts rather than merely natural-language
queries. Chen et al.~\cite{Chen:TOSEM:2026} characterize prompts as
first-class artifacts in prompt-enabled systems and propose
``promptware engineering'' as a systematic methodology for prompt
requirements, design, testing, evolution, deployment, and monitoring.
Villamizar et al.~\cite{Villamizar:PROFES:2025} similarly emphasize
prompt management, reuse, traceability, and maintenance, while Liang
et al.~\cite{Liang:FSE:2025} provide qualitative evidence that
developers engage in iterative ``prompt programming,'' treating
prompts as programmable artifacts that require refinement, testing,
and evaluation.

Collectively, these studies establish that prompt formulation is an
important developer-facing artifact that shapes LLM behavior and
software-engineering outcomes. However, prior work primarily focuses
on conversational effectiveness, prompting practices, code quality, or
prompt management. In contrast, our study examines how distinct
prompt-structure dimensions influence successive stages of
AI-assisted pull request workflows, including actionable code
generation, code adoption, and integration depth.

\subsection{Pull Request Evaluation and Developer-Facing Artifacts}

Prior work shows that successful software collaboration depends on how
effectively technical information is communicated to developers and
reviewers. Studies of bug-report quality emphasize the importance of
contextual information such as reproduction steps, stack traces, test
cases, and environment details that help developers localize and
understand a problem~\cite{bettenburg2008goodbugreport,
zimmermann2010goodbugreport}. Similarly, research on pull request
evaluation shows that contribution acceptance depends on both
technical and social signals, including change quality, contributor
experience, review discussion, and repository context~\cite{tsay2014githubfactors,
gousios2016contributorperspective}. These findings motivate our
\textit{Context} and \textit{Specificity} dimensions, which capture
technical grounding and the communication of a bounded, actionable
task.

Prior work also highlights the importance of evidence that enables
developers to assess whether a change is correct or acceptable.
External references help document and justify pull request
changes~\cite{zampetti2017externalreferences}, while continuous
integration and testing provide automated signals for evaluating
incoming contributions~\cite{vasilescu2015ciquality}. These studies
motivate our \textit{Verification} dimension, which captures whether
a prompt provides information that supports correctness assessment,
such as tests, expected behavior, examples, constraints, or
references to specifications.

Building on this literature, we treat prompts as developer-facing
artifacts that communicate context, intent, and evaluability cues in
much the same way as bug reports and pull request descriptions,
thereby influencing downstream development and review decisions.

\subsection{LLM-Assisted Annotation and Empirical SE Methodology}

LLMs are increasingly used not only as software-engineering tools,
but also as instruments for empirical research, including coding,
labeling, synthesis, and artifact assessment. Recent methodological
guidelines for empirical SE studies involving LLMs emphasize that
researchers should explicitly report the role of the LLM, model
configuration, prompts, interaction logs, validation procedures, and
limitations, particularly when LLMs are used as annotators or
judges~\cite{baltes2025llmguidelines}. These recommendations are
especially relevant because LLM outputs are affected by model
non-determinism, evolving model versions, prompt sensitivity, and
context-dependent interpretation. Broader work on LLM-based data
annotation similarly shows that LLMs can reduce annotation cost
and scale empirical analysis, but that their labels require careful
assessment before being used as ground truth~\cite{tan-etal-2024-large}.

Recent human--LLM annotation studies further caution against
fully replacing human judgment with automated labels. Wang et
al.~\cite{wang:CHI:2024} propose a collaborative
annotation framework in which LLMs first generate labels and
explanations, a verifier identifies potentially unreliable labels, and
humans re-annotate low-confidence cases. 
Similarly, Takehi et al.~\cite{Takehi:SIGIR:2025} show that LLM
judgments can assist large-scale relevance assessments, but that their
effectiveness varies across assessment contexts. They therefore
advocate selective human involvement to calibrate and verify LLM
predictions where reliability is uncertain. 
Xu et al.~\cite{Xu:LAK26:2026} further argue
that aggregate agreement metrics can obscure important differences
between task-inherent ambiguity and model-driven errors,
especially for ordinal and interpretive annotation tasks. These
studies motivate our treatment of prompt-quality annotation as a
methodological research question rather than a preprocessing step.
Accordingly, we evaluate human--LLM agreement across prompt
dimensions and workflow outcome classes, and derive a
class-aware hybrid annotation policy that uses LLM labels only
where they are sufficiently aligned with human judgment while
retaining human annotation where reliability concerns remain or where
additional human verification is warranted.

Taken together, prior work establishes that prompt construction
influences LLM behavior and that prompts increasingly function as
developer-facing artifacts within collaborative AI-assisted
workflows. Existing studies, however, primarily focus on
conversational effectiveness, prompt-development practices, code
quality, or artifact integration outcomes. In contrast, our study
examines how distinct prompt-structure dimensions influence
successive stages of AI-assisted pull request workflows and evaluates
whether these dimensions can be measured reliably through
LLM-assisted annotation.

\section{Threats to Validity}
\label{sec:threats}

\paragraph{Construct Validity.}
Our measurement of prompt structure relies on the operationalization of
Context, Specificity, and Verification. Although these dimensions are
grounded in prior software-engineering and prompt-engineering research,
they may not capture all characteristics of effective prompting or
developer intent. Prompt quality in practice may additionally depend on
factors such as conversational strategy, domain knowledge, reasoning
style, or broader interaction history that are not explicitly modeled by
our rubric. To mitigate subjectivity, we employed a structured
codebook, conducted blinded human annotation on a validation subset,
and evaluated agreement with LLM-assisted annotation using ordinal
agreement metrics, error analysis, and bias assessment. Because
annotation reliability varied across dimensions and workflow outcome
classes, we derived a class-aware hybrid annotation policy in which
dimensions exhibiting instability or systematic bias remained
human-scored. The final analytical dataset used in RQ2 was constructed
using this empirically validated policy rather than a uniform automated
annotation strategy, reducing the risk that downstream findings are
driven by annotation artifacts.   
Agreement estimates for some outcome-class subsets (particularly CL,
and to a lesser extent NE for Verification) should nevertheless be
interpreted cautiously because small validation sample sizes, skewed
score distributions, and prevalence effects can produce unstable
agreement statistics, including negative weighted kappa estimates
despite relatively high observed agreement.
We additionally provide the annotation prompts, codebook, and replication
artifacts to support transparency and reproducibility.

\paragraph{Internal Validity.}
The study design is observational and does not establish causal
relationships between prompt structure and pull request outcomes.
Although we control for factors such as pull request size,
contributor experience, repository context, and prompt length,
unobserved confounding variables may still influence the observed
relationships. In addition, the stage-based workflow framework
represents an analytical abstraction of collaborative development
behavior rather than a strictly sequential process, and some workflow
stages may partially overlap in practice.

A further threat concerns conversational completeness. The shared
ChatGPT conversations analyzed in this study may not fully capture all
developer--LLM interactions associated with a pull request. Developers
may omit intermediate prompts, continue conversations outside publicly
shared artifacts, or substantially revise generated outputs before
integration. Consequently, the observed prompts may represent partial
views of broader iterative workflows. 
To mitigate these risks, we conducted robustness checks, sensitivity
analyses, hold-out stability tests, and alternative model
specifications. These analyses preserved the principal
stage-dependent relationships, although some effects became less
stable in smaller restricted samples, particularly for Gate~2
integration-depth models estimated on PA-only subsets.

\paragraph{External Validity.}
Our dataset is derived from self-admitted ChatGPT usage (SACU) cases in
public GitHub pull request workflows. Consequently, the findings may
not generalize to undisclosed AI usage, private development settings,
or interactions involving other LLMs and coding agents. As observed in
prior work on self-admitted GenAI usage in open-source software
development~\cite{Xiao:TSE:2026}, developers who publicly share
AI conversations may differ behaviorally from those who do not
disclose AI assistance, introducing potential survivorship and
disclosure bias into the observed workflow patterns. Furthermore, the
dataset reflects interactions collected during earlier generations of
ChatGPT, and newer models may respond differently to the same prompts.
However, our focus is on how prompt structure shapes collaborative
workflow behavior, including code generation, adoption, and
integration, which is less sensitive to specific model versions than
raw model capability. Future work should examine whether these
workflow patterns persist across newer LLM ecosystems and alternative
collaborative development settings.

\paragraph{Conclusion Validity.}
Our conclusions are based on statistical modeling and agreement
analysis, both of which may be affected by sample size, class
imbalance, model assumptions, and measurement error. In addition, the
workflow outcome classes used throughout the analysis are inherited
from the manually validated PatchTrack dataset, and residual
classification ambiguity may still affect some downstream analyses,
particularly for partially integrated or substantially adapted patches.
We mitigate these risks through model diagnostics,
repository-clustered robustness specifications, sensitivity analyses,
hold-out stability checks, and alternative operationalizations of
workflow outcomes.
Across these analyses, the principal stage-dependent relationships
remained qualitatively stable, with \textit{Specificity} and
\textit{Context} supporting code generation, \textit{Verification}
remaining most strongly associated with code adoption, and
\textit{Context} continuing to exhibit the strongest relationship
with integration depth.
Additionally, the qualitative analysis was intended to contextualize
quantitative findings rather than support standalone thematic claims.
Consequently, the examples should be interpreted as illustrative of
recurring patterns rather than representative estimates of their
prevalence, reducing the risk of overgeneralization.

\subsection{Limitations and Future Work}

Although the proposed stage-based framework provides a useful
abstraction for modeling AI-assisted pull request workflows,
real-world collaborative development often involves overlapping,
iterative, and conversationally dependent behaviors that may not be
fully captured by discrete analytical stages. Similarly, our
operationalization of prompt structure through Context, Specificity,
and Verification intentionally emphasizes interpretable developer-facing
dimensions, but may not capture finer-grained conversational or semantic
characteristics that influence LLM-assisted development outcomes.

Future work should investigate how prompt-aware tooling, review support
systems, and developer education practices can leverage workflow-aware
prompt characteristics to improve collaborative AI-assisted software
engineering. Additional research could also examine richer
conversation-level representations, longitudinal developer--LLM
interaction patterns, and the role of emerging agent-based coding
systems in shaping collaborative development workflows.

\section{Ethical Implications}

This study analyzes publicly accessible GitHub pull requests and
developer-shared ChatGPT conversations involving self-admitted AI
usage. Although these artifacts are publicly available, we recognize
the importance of responsible handling of developer-generated content.
To minimize privacy risks, we restrict our analysis to publicly shared
artifacts, avoid collecting private or personally sensitive
information, and present illustrative examples only when necessary to
support empirical findings. We additionally provide a public
replication package containing datasets, analysis scripts, annotation
artifacts, and supplementary documentation to support transparency,
reproducibility, and responsible reuse of the research
artifacts~\cite{sserunjogi_2026_20451326}.

\section{Conclusion}
\label{sec:conclusion}

This paper investigated how prompt structure relates to progression
through AI-assisted pull request workflows. Using a manually validated
dataset of 265 pull request-linked developer--ChatGPT interactions, we
examined how three prompt dimensions, \textit{Context},
\textit{Specificity}, and \textit{Verification}, relate to actionable
code generation, code adoption, integration depth, and broader
pull request workflow outcomes.

Our findings show that prompt characteristics exert distinct
stage-dependent associations across collaborative software
engineering workflows. Specifically, \textit{Specificity} and \textit{Context}
are most strongly associated with actionable
code generation, \textit{Verification} emerges as the primary
predictor of code adoption, and \textit{Context} again becomes the
dominant factor for deeper integration. These results highlight the
importance of contextual grounding, evaluability, and implementation
alignment throughout AI-assisted pull request workflows.

The study further demonstrates that reliable
operationalization of prompt structure requires adaptive hybrid
human--LLM annotation strategies rather than uniform automation, as
annotation reliability varies substantially across prompt dimensions
and workflow contexts. More broadly, the results show that the
reliability of LLM-assisted annotation depends both on the construct
being measured and the context in which it is applied, highlighting
the need for construct-aware and context-aware validation in future
empirical studies.

Overall, the results suggest that effective AI-assisted software
engineering depends not only on generating plausible code, but also on
how developers construct prompts that support subsequent evaluation,
adoption, adaptation, and integration activities. Viewed collectively,
the findings position prompt construction as an upstream factor in
AI-assisted software-development workflows, influencing not only code
generation but also downstream collaborative outcomes within pull
request processes. 
We hope this work motivates future research on
workflow-aware prompt engineering, context-aware AI-assisted
development tools, scalable methodologies for studying
developer--LLM interactions, and human-centered evaluation
frameworks for collaborative software engineering with LLMs.

\section*{Declarations}
\subsection*{Funding}
This research was supported by NSF Grant Award No.~\#2519136.

\subsection*{Ethics approval}
Not applicable
\subsection*{Informed consent}
Not applicable
\subsection*{Author contribution}
\textbf{John Businge}: Conceptualization, supervision, and review and editing of the manuscript.\\
\textbf{Richard Sserunjogi}: Data collection, analysis,  writing of the original draft.\\
\textbf{Daniel Ogenrwot}: Data analysis and writing of the original draft.

\subsection*{Data availability}
The datasets, analysis scripts, and supplementary documentation are publicly available at \href{https://zenodo.org/records/20451325}{https://zenodo.org/records/20451325}.

\subsection*{Conflict of interest}
The authors declare that there are no known competing financial interests or personal relationships that could have appeared to influence the work reported in this paper.

\subsection*{Clinical trial number in the manuscript}
Not applicable

\bibliography{biblio}

\end{document}